\documentclass[oneside,a4paper]{article}
\usepackage{hyperref}
\usepackage{amsmath,bm}
\usepackage{amssymb}
\usepackage[mathscr]{euscript}
\usepackage{mathrsfs}
\usepackage{amsfonts}
\usepackage{ntheorem}
\usepackage{geometry}
\usepackage{bbm}
\usepackage{titlesec}
\usepackage[x11names]{xcolor}
\usepackage{float}
\usepackage{graphicx}
\usepackage{braket}
\usepackage{verbatim}
\usepackage{subfigure}
\usepackage{tcolorbox}
\usepackage[capitalize]{cleveref} % \cref
\usepackage{authblk}
\usepackage[misc]{ifsym}
\usepackage{hyperref}

\DeclareRobustCommand{\rchi}{{\mathpalette\irchi\relax}}
\newcommand{\irchi}[2]{\raisebox{\depth}{$#1\chi$}}

\hypersetup{
    colorlinks=true,
    linkcolor=black,      % 链接引用颜色，包括章节和方程
    citecolor=blue,     % 文献引用颜色
}

\titleformat{\chapter}[hang]{\normalfont\Huge\bfseries}{\thechapter}{1em}{}

\tcbuselibrary{skins, breakable, theorems}

\theoremstyle{break}
\theoremseparator{:}

\theorembodyfont{\upshape}

\newtheorem{de}{\textbf{Definition}}

\newtheorem{theorem}{\textbf{Theorem}}
\newtheorem*{pro}{\textbf{Proof}}
\newtheorem{lemma}{\textbf{Lemma}}

\newtheorem{cor}{\textbf{Corollary}}

\newtheorem{propo}{\textbf{Proposition}}

\allowdisplaybreaks[4]

\newcommand{\bff}{\mathbf}

\newcommand{\ct}{\mathcal{T}}

\newcommand{\cs}{\mathcal{S}}

\newcommand{\cx}{\mathcal{X}}
\newcommand{\cy}{\mathcal{Y}}

\newcommand{\ca}{\mathcal{A}}
\newcommand{\cb}{\mathcal{B}}
\newcommand{\cc}{\mathcal{C}}
\newcommand{\cd}{\mathcal{D}}

\newcommand{\cm}{\mathcal{M}}

\newcommand{\cu}{\mathcal{U}}
\newcommand{\cv}{\mathcal{V}}
\newcommand{\cw}{\mathcal{W}}

\newcommand{\cg}{\mathcal{G}}
\newcommand{\ccp}{\mathcal{CP}}
\newcommand{\cf}{\mathcal{F}}

\newcommand{\ce}{\mathcal{E}}

\newcommand{\supp}{\mathrm{supp}}

\newcommand{\spann}{\operatorname{span}}

\newcommand{\wt}{\operatorname{wt}}
\newcommand{\rk}{\operatorname{rank}}

\newcommand{\srk}{\operatorname{srk}}

\newcommand{\pr}{\mathop{\mathbf{Pr}}\limits}

\newcommand{\stirling}[2]{\genfrac{[}{]}{0pt}{}{#1}{#2}}

\title{On the List-Decodability of Random (Linear) Sum-Rank Metric Codes}

\author[1]{Yang Liu\thanks{Corresponding author: Yang Liu \{ge94vib@mytum.de\}}}
\author[2]{Anna Baumeister}
\author[2]{Antonia Wachter-Zeh}

\affil[1]{Technical University of Munich}
\affil[2]{Chair of Coding and Cryptography, Technical University of Munich}

\begin{document}

% Hello Anna, I made a small edit today without changing the statement you had already edited yesterday.

\maketitle
\begin{abstract}
In this paper, we establish the list-decoding capacity theorem for sum-rank metric codes. This theorem implies the list-decodability theorem for random general sum-rank metric codes: Any random general sum-rank metric code with a rate not exceeding the list-decoding capacity is $\left(\rho,O\left(1/\epsilon\right)\right)$-list-decodable with high probability, where $\rho\in\left(0,1\right)$ represents the error fraction and $\epsilon>0$ is referred to as the capacity gap. For random $\mathbb{F}_q$-linear sum-rank metric codes by using the same proof approach we demonstrate that any random $\mathbb{F}_q$-linear sum-rank metric code with a rate not exceeding the list-decoding capacity is $\left(\rho,\exp\left(O\left(1/\epsilon\right)\right)\right)$-list-decodable with high probability, where the list size is exponential at this stage due to the high correlation among codewords in linear codes. To achieve an exponential improvement on the list size, we prove a limited correlation property between sum-rank metric balls and $\mathbb{F}_q$-subspaces. Ultimately, we establish the list-decodability theorem for random $\mathbb{F}_q$-linear sum-rank metric codes: Any random $\mathbb{F}_q$-linear sum-rank metric code with rate not exceeding the list decoding capacity is $\left(\rho, O\left(1/\epsilon\right)\right)$-list-decodable with high probability. For the proof of the list-decodability theorem of random $\mathbb{F}_q$-linear sum-rank metric codes our proof idea is inspired by and aligns with that provided in the works \cite{Gur2010,Din2014,Gur2017} where the authors proved the list-decodability theorems for random $\mathbb{F}_q$-linear Hamming metric codes and random $\mathbb{F}_q$-linear rank metric codes, respectively.
\end{abstract}

\section{Introduction}
\textit{List-decoding} is a fundamental notion in coding theory, which was proposed independently by Elias and Wozencraft \cite{Eli1957,Woz1958}. In contrast to traditional unique decoding, which outputs a single valid codeword within a certain maximal distance from the received word, list decoding outputs all codewords within a certain radius that usually exceeds the unique decoding radius. The three important parameters involved in the research on list decodability are: \textit{code rate} $R$, \textit{error fraction} $\rho$, and \textit{list size} $L$, and one of the key points of the research is the interaction between these three parameters.

The sum-rank metric is a comparatively new metric, first introduced in the context of multishot network coding \cite{Nbr2010}. In this metric, a vector of length $n$ over an extension field $\mathbb{F}_{q^m}$ is partitioned into $\ell$ blocks, and its weight is defined as the sum of the ranks of each block (where the rank of a block is computed from its matrix representation over $\mathbb{F}_q$). Thus, the sum-rank metric can be viewed as a generalization of the Hamming metric ($n$ blocks of length one) and the rank metric (one block of length $n$).
In the past $45$ years, the list-decodability of random ($\mathbb{F}_q$-linear) Hamming metric codes \cite{Zya1981,Eli1991,Gur2002,Gur2010,Li2018} and random ($\mathbb{F}_q$-linear) rank metric codes \cite{Din2014,Gur2017,Res2019} has been studied. In this paper, we investigate whether random sum-rank metric codes exhibit list decodability properties similar to those of random ($\mathbb{F}_q$-linear) Hamming metric codes and random ($\mathbb{F}_q$-linear) rank metric codes.

\subsection{Related Works}
We now provide a summary of previous studies on the list-decodability of random ($\mathbb{F}_q$-linear) Hamming metric codes and random ($\mathbb{F}_q$-linear) rank metric codes.
\begin{enumerate}
    \item List-decodability of random ($\mathbb{F}_q$-linear) Hamming metric codes. The study of the list-decodability of random $\mathbb{F}_q$-linear Hamming metric codes can be traced back to the following conclusion about the list-decodability of random $\mathbb{F}_2$-linear Hamming metric codes proved by Zyablov and Pinsker in 1981 \cite{Zya1981}: A random $\mathbb{F}_2$-linear Hamming metric code of rate $1-H_2\left(\rho\right)-\epsilon$ is $(\rho,2^{O\left(1/\epsilon\right)})$-list-decodable with high probability. This conclusion can be directly extended to random $\mathbb{F}_q$-linear Hamming metric codes: A random $\mathbb{F}_q$-linear Hamming metric code of rate $1-H_q\left(\rho\right)-\epsilon$ is $(\rho,q^{O\left(1/\epsilon\right)})$-list-decodable with high probability, where the function $H_q\left(\cdot\right)$ is the $q$-ary entropy function. The quantity $1-H_{q}\left(\rho\right)$ is said to be the \textit{list-decoding capacity of Hamming metric codes}. 10 years later, in \cite{Eli1991} Elias investigated whether the list size for random $\mathbb{F}_q$-linear Hamming metric codes can achieve a list size comparable to the optimal list size $O\left(1/\epsilon\right)$ of random general Hamming metric codes \cite{Gur2010}. In \cite{Gur2002,Li2018} Guruswami et al. and Li et al. showed that with high probability a random $\mathbb{F}_2$-linear Hamming metric code of rate $1-H_{2}\left(\rho\right)-\epsilon$ is $\left(\rho,O\left(1/\epsilon\right)\right)$-list-decodable \cite{Res2019}. The next breakthrough involving larger alphabets occurred in 2010. Guruswami et al. in \cite{Gur2010} showed the following result, addressed the aforementioned unresolved issue regarding the list-decodability of random $\mathbb{F}_q$-linear Hamming metric codes for $q\ge 2$ by showing a limited correlation property between Hamming balls and $\mathbb{F}_q$-subspaces: Any random $\mathbb{F}_q$-linear Hamming metric code with rate \textit{not} exceeding the list-decoding capacity is $\left(\rho,O\left(1/\epsilon\right)\right)$-list-decodable with high probability.
    \item List-decodability of random ($\mathbb{F}_q$-linear) rank metric codes. Ding in \cite{Din2014} first started to study the list-decodability of random ($\mathbb{F}_q$-linear) rank metric codes. First, the author showed the so-called list-decoding capacity theorem for rank metric codes \cite{Res2019} which implies the following list-decodability theorem for random general rank metric codes: Any random general rank metric code in $\mathbb{F}_{q}^{m\times n}$ with rate $1-(\rho+\rho b-\rho^2b)-\epsilon$, where $b:=n/m$, is $\left(\rho,O\left(1/\epsilon\right)\right)$-list-decodable with high probability. The quantity $1-(\rho+\rho b-\rho^2b)$ is referred to as the \textit{list-decoding capacity of rank metric codes}. He also provided a result for random $\mathbb{F}_q$-linear rank metric codes: Any random $\mathbb{F}_q$-linear rank metric code with rate \textit{not} exceeding the list-decoding capacity is $\left(\rho,\exp\left(O\left(1/\epsilon\right)\right)\right)$-list-decodable with high probability. The subsequent work \cite{Gur2017} achieved an exponential improvement on the list size by showing a limited correlation property between rank metric balls and $\mathbb{F}_q$-subspaces, and the proof framework is consistent with that in \cite{Gur2010,Din2014}: Any random $\mathbb{F}_q$-linear rank metric code with rate \textit{not} exceeding the list-decoding capacity is $\left(\rho,O\left(1/\epsilon\right)\right)$-list-decodable with high probability.
\end{enumerate}

\subsection{Contributions and Organization}
We first provide the necessary background knowledge with regard to combinatorics and list decoding in Section \ref{sec:1}. Subsequently, we introduce a new tool called decomposable subspaces and its fundamental properties in Section \ref{sec:2}, which will assist in proving the main results. Proofs for these two sections are collected in the Appendix to improve readability.
The first main result is the \textit{list}-\textit{decoding capacity theorem for sum-rank metric codes}, which also implies the \textit{list-decodability theorem for random general sum-rank metric codes} in Section \ref{sec:3}. The proof technique is consistent with that in \cite[Theorem III.1]{Din2014} and \cite[Theorem 2.4.7]{Res2019}. Subsequently, by employing the same proof techniques, we prove a trivial version of the list-decodability theorem for random $\mathbb{F}_q$-linear sum-rank metric codes. Based on the new construction of \textit{decomposable subspaces}, we first establish the so-called dimension lemma (cf. Lemma \ref{lem:7}) and Theorem \ref{th:4}, which serves as a key result for completing the list-decodability theorem for random $\mathbb{F}_q$-linear sum-rank metric codes. Following the approach outlined in \cite{Gur2017, Res2019} we prove the limited correlation property between sum-rank metric balls and $\mathbb{F}_q$-subspaces (Lemma \ref{lem:6}). Both Theorem \ref{th:4} and Lemma \ref{lem:6} generalize the corresponding results of Hamming metric \cite{Gur2010} and rank metric \cite{Gur2017}. Finally, leveraging the same proof framework as in \cite{Gur2010, Din2014, Gur2017}, we complete the proof of the list-decodability theorem for random $\mathbb{F}_q$-linear sum-rank metric codes. Finally, we summarize our findings and offer a brief outlook on future works in Section \ref{sec:4}.

\section{Preliminaries}\label{sec:1}
Throughout this paper, let $q$ be some prime power and $\mathbb{F}_q$ be a finite field. We use $\left[a:b\right]$ to represent the set of natural numbers $\left\{a,a+1,\ldots, b\right\}$. Uppercase Latin letters $A,B,\ldots$ are used to represent matrices, (random) variables, and other non-algebraic structures; calligraphic uppercase letters $\mathcal{A},\mathcal{B},\ldots$ to denote sets, vector spaces, and other algebraic structures; bold uppercase letters $\bm{A},\bm{B},\ldots$, bold lowercase letters $\bm{a},\bm{b},\ldots$, and bold lowercase Greek letters $\bm{\alpha},\bm{\beta},\ldots$ are used to represent ordered tuples of non-algebraic structures. The lower case Latin letters $a,b,\ldots$ and lower case Greek letters $\alpha,\beta,\ldots$ are also used to represent scalars. The product of $\ell$ sets or $\ell$ algebraic structures $\ca_1,\ca_2,\ldots,\ca_{\ell}$ will be denoted as $\Pi_{i}^{\ell}\ca_i:=\ca_1\times\ca_2\times\ldots\times\ca_{\ell}$. In particular, if $\ca_i=\ca$ for all $i\in\left[1:\ell\right]$, we denote $\Pi_{i}^{\ell}\ca_i$ as $\ca^{\ell}$. If $\ca_i=\mathbb{F}_q^{m_i\times{\eta}_i}$ for all $i\in\left[1:\ell\right]$, where $\mathbb{F}_q^{m_i\times\eta_i}$ is the $\mathbb{F}_q$-vector space of all $\left(m_i\times \eta_i\right)$-matrices over $\mathbb{F}_q$, we denote $\Pi_i^{\ell}\mathbb{F}_q^{m_i\times\eta_i}$ simply as $\cm_i^{\ell}$, and for the special case that $m_i=m$ and $\eta_i=\eta$ for all $i\in\left[1:\ell\right]$, we write $\cm_i^{\ell}$ as $\cm^{\ell}$. We write $\ca\subseteq\cb$ as usual if $\ca$ is a subset of $\cb$, and $\cu\le_{\mathbb{F}_q}\cv$ if $\cu$ is an $\mathbb{F}_q$-subspace of $\mathbb{F}_q$-vector space $\cv$ to emphasize its linear structure. Unless otherwise specified, we adopt the following conventions: We omit all unnecessary occurrences of $\mathbb{F}_q$. For instance, $\mathbb{F}_q$-subspaces will be referred to as subspaces, $\mathbb{F}_q$-linear independence will be referred to as independence, etc. When $\mathbb{F}_q$ appears as a subscript for certain operators, it will also be omitted. For example, $\le_{\mathbb{F}_q}$ will be abbreviated as $\le$ (although this conflicts with the symbol for the order relation on real numbers, readers can easily infer the actual meaning of $\le$ from the objects on either side), $\dim_{\mathbb{F}_q}$, i.e., the dimension with respect to $\mathbb{F}_q$, will be uniformly abbreviated as $\dim$, and $\spann_{\mathbb{F}_q}$, i.e., the linear span operator with respect to $\mathbb{F}_q$, will be abbreviated as $\spann$.

In this paper, many conclusions and proofs rely on the concept of \textit{integer partitions}, which is a fundamental concept in combinatorics, referring to the decomposition of a positive integer into a sum of several positive integers. Based on whether the order of the parts matters, integer partitions can be classified into \textit{ordered partitions} (also known as \textit{compositions}) and \textit{unordered partitions}. Furthermore, based on whether constraints are imposed, partitions can be categorized as \textit{constrained} or \textit{unconstrained}. A type particularly relevant to our purposes is the \textit{constrained ordered partition}. For an in-depth discussion on integer partitions, we recommend the excellent references \cite{And1984,Gra1991,Wil2006,Sta2011}.
\begin{de}[Constrained Ordered $\ell$-Partition of a Positive Integer \cite{And1984}]
Let $w$ be a positive integer. A \textit{constrained ordered} $\ell$-\textit{partition} of $w$ is an ordered $\ell$-tuple $\bm{w}:=\left(w_1,w_2,\ldots,w_{\ell}\right)$ of non-negative integers such that $0\le a_i\le w_i\le b_i$ for all $i\in\left[1:\ell\right]$, and $w=w_1+w_2+\ldots+w_{\ell}$. The set of all constrained ordered $\ell$-partitions of $w$ will be denoted as $\ccp_{\ell}\left(w\right)$.
\end{de}

By \cite{Rat2008,Ott2023} we have:
\begin{equation*}
|\ccp_{\ell}\left(w\right)|\le\binom{w+\ell-1}{\ell-1}.
\end{equation*}

\noindent Two concepts closely related to this paper are the \textit{Grassmannian} and the \textit{Gaussian binomial coefficient}.
\begin{de}[Grassmannian, Gaussian Binomial Coefficient]
Let $\mathbb{F}$ be a field. For a fixed $\mathbb{F}$-vector space $\cv$, the \textit{Grassmannian} $\cg_{\mathbb{F}}\left(k,\cv\right)$ represents the set of all $k$-dimensional $\mathbb{F}$-subspaces of $\cv$. For $\mathbb{F}=\mathbb{F}_q$ and $\dim\left(\cv\right)=\eta$, the size of $\cg_{\mathbb{F}_q}\left(k,\cv\right)$ can be determined by the \textit{Gaussian binomial coefficient}, or $q$-\textit{binomial coefficient} \cite{Sta2011}:
\begin{equation}\label{eq:1}
\left|\cg_{\mathbb{F}_q}\left(k,\cv\right)\right|=:\begin{bmatrix}
\eta \\ k
\end{bmatrix}_q:=\left\{
\begin{array}{ll}
\displaystyle{\prod_{i=0}^{k-1}\frac{q^{\eta}-q^i}{q^k-q^i},} & \displaystyle{k\le\eta,} \vspace{1ex}\\
\displaystyle{0,} & \displaystyle{\mathrm{otherwise}.}
\end{array}
\right.
\end{equation}
\end{de}

Lemma \ref{lem:1} provides two useful basic properties of the $q$-binomial coefficient.
\begin{lemma}[\cite{Koe2007,Gru2023}]
Let $d$, $k$, and $\eta$ be integers with $d\le k\le \eta$. Then:
\begin{enumerate}
\item $q^{\left(\eta-k\right)k}\le\stirling{\eta}{k}_q\le K_q^{-1}q^{\left(\eta-k\right)k}$, where $K_q^{-1}:=\prod_{i=1}^{\infty}(1-q^{-i})^{-1}<4$.
\item The $q$-binomial coefficient $\stirling{\eta-d}{k-d}_q$ is the number of $k$-dimensional subspaces of an $\eta$-dimensional $\mathbb{F}_q$-vector space $\cv$, which contains a fixed $d$-dimensional subspace of $\cv$.
\end{enumerate}
\label{lem:1}
\end{lemma}

The \textit{sum-rank metric} was first introduced in \cite{Nbr2010} which covers the Hamming and rank metrics. There are two equivalent definitions of the sum-rank metric, one via vectors and one via matrices. In this paper, we focus exclusively on the \textit{matrix sum-rank metric}, henceforth simply referred to as sum-rank metric. The sum-rank metric is established based on the concept of \textit{sum-rank weight}.
\begin{de}[Sum-rank weight]
The sum-rank weight of  any $\bm{X}:=\left(X_1,X_2,\ldots,X_{\ell}\right) \in \cm^{\ell}$ is defined as: 
\begin{equation*}
\wt_{\srk,\ell}\left(\bm{X}\right):=\sum_{i=1}^{\ell}\rk\left(X_i\right),
\end{equation*}
    i.e., the weight of  $\bm{X}$ is the sum of the ranks of its $\ell$ blocks.
\end{de}

The sum-rank distance between $\bm{X}$ and $\bm{Y}$ is the sum-rank weight of their difference:
\[ d_{\srk,\ell}\left(\bm{X},\bm{Y}\right):=\wt_{\srk,\ell}\left(\bm{X}-\bm{Y}\right).\]
Finally, the sum-rank distance defines the \textit{sum-rank metric} over $\cm^{\ell}$.

As in any metric space, we have the concept of sum-rank metric spheres and balls.

\begin{de}[Sum-rank metric spheres and balls]
Let $\bm{X}\in\cm^{\ell}$ be fixed and $r\in\mathbb{Z}^+$. Then the \textit{sum-rank metric sphere} with center $\bm{X}$ and radius $r$ is defined as the set of all vectors $\bm{Y}$ with distance exactly $r$ to $\bm{X}$.
\begin{equation*}
\cs_{\srk,\ell}(\bm{X},r):=\{\bm{Y}\in\cm^{\ell}:d_{\srk,\ell}\left(\bm{X},\bm{Y}\right)=r\}.
\end{equation*} 
The \textit{sum-rank metric ball} with center at $\bm{X}$ and radius $r$ is the set of vectors $\bm{Y}$ with distance \textbf{at most} $r$ from $\bm{X}$:
\begin{equation*}
\cb_{\srk,\ell}(\bm{X},r):=\{\bm{Y}\in\cm^{\ell}:d_{\srk,\ell}\left(\bm{X},\bm{Y}\right)\le r\}.
\end{equation*}
%Clearly, volumes (cardinalities of sets) of $\cs_{\srk,\ell}\left(\bm{X},r\right)$ and $\cb_{\srk,\ell}\left(\bm{X},r\right)$ depend \textit{only} on the radius $r$.
\end{de} 
For any $\bm{X}$ we have:
\begin{equation}\label{eq:SRBallsize}
\left|\cb_{\srk,\ell}\left(\bm{X},r\right)\right|=\sum_{s=0}^{r}\left|\cs_{\srk,\ell}\left(\bm{X},s\right)\right|.
\end{equation}

\noindent It is easy to show that the sum-rank metric is \textit{translation and scaling invariant} with respect to its two parameters.
\begin{propo}\label{prop:2}
Let $\bm{X},\bm{Y},\bm{Z}\in\cm_{i}^{\ell}$ and $\alpha\in\mathbb{F}_q\setminus\{0_{\mathbb{F}_q}\}$. Then the sum-rank metric $d_{\srk,\ell}$ is both \textit{translation} and \textit{scaling} invariant with respect to its two parameters, that is $d_{\srk,\ell}\left(\bm{X}\pm\bm{Z},\bm{Y}\pm\bm{Z}\right)=d_{\srk,\ell}\left(\bm{X},\bm{Y}\right)$ and $d_{\srk,\ell}\left(\alpha\bm{X},\alpha\bm{Y}\right)=d_{\srk,\ell}\left(\bm{X},\bm{Y}\right)$.
\end{propo}

As a direct consequence of Proposition \ref{prop:2} we have the following corollary.
\begin{cor}\label{cor:1}
Let $\bm{X},\bm{Y}\in\cm_i^{\ell}$, $\alpha\in\mathbb{F}_q\setminus\{0_{\mathbb{F}_q}\}$, and $r\in\mathbb{N}_0$. Then $\bm{X}\in\cb_{\srk,\ell}\left(\bm{0},r\right)$ if and only if $\bm{X}\pm\bm{Y}\in\cb_{\srk,\ell}\left(\pm\bm{Y},r\right)$, and $\bm{X}\in\cb_{\srk,\ell}\left(\bm{0},r\right)$ if and only if $\alpha\bm{X}\in\cb_{\srk,\ell}\left(\bm{0},r\right)$.
\end{cor}

For the remainder of this paper we will consider only the special case where all blocks have the same size, i.e., $m_i=m$ and $\eta_i=\eta$ for all $i\in\left[1:\ell\right]$, that is, the $\mathbb{F}_q$-vector space $\cm^{\ell}$. 

Authors in \cite{Ott2023} have proven that:
\begin{equation*}
|\cs_{\srk,\ell}\left(\bm{0},r\right)|=\left\{
\begin{array}{ll}
\displaystyle{1,} & \displaystyle{r=0,} \vspace{1ex}\\
\displaystyle{\sum_{\bm{r}\in\ccp_{\ell}\left(r\right)}
\prod_{i=1}^{\ell}\prod_{j=0}^{r_i-1}\frac{\left(q^{m}-q^j\right)\left(q^{\eta}-q^j\right)}{q^{r_i}-q^j},} & \displaystyle{r\ge 1,}
\end{array}
\right.
\end{equation*}
where $r_i$ is the $i$-th coordinate of $\bm{r}$ for each $i\in\left[1:\ell\right]$. Thus by \cref{eq:SRBallsize}:
\begin{equation*}
|\cb_{\srk,\ell}\left(\bm{0},r\right)|=\sum_{j=0}^{r}\sum_{\bm{j}\in\ccp_{\ell}\left(j\right)}
\prod_{i=1}^{\ell}\prod_{k=0}^{j_i-1}\frac{\left(q^{m}-q^k\right)\left(q^{\eta}-q^k\right)}{q^{j_i}-q^k},
\end{equation*}
where $j_i$ is the $i$-th coordinate of $\bm{j}$ for each $i\in\left[1:\ell\right]$. Moreover, authors in \cite{Puc2020,Ott2021} have provided lower and upper bounds on $|\cs_{\srk,\ell}\left(\bm{0},r\right)|$, respectively:
\begin{equation}\label{eq:4}
K_q^{\ell}q^{\left(m+\eta-r/\ell\right)r-\ell/4}\le|\cs_{\srk,\ell}\left(\bm{0},r\right)|\le K_{q}^{-\ell}\binom{\ell+r-1}{r}q^{\left(m+\eta-r/\ell\right)r}.
\end{equation}

We now provide a lower bound and an upper bound for the volume of the sum-rank metric balls with some radius $r$. These bounds are crucial for many proofs in the subsequent results.
\begin{propo}\label{prop:1}
The volume of the sum-rank metric ball $\cb_{\srk,\ell}\left(\bm{0},r\right)$ for any $r\in\left[0:\ell\min\left\{m,\eta\right\}\right]$ can be bounded from below and above by:
\begin{equation*}
K_{q}^{\ell}q^{\left(m+\eta-r/\ell\right)r-\ell/4}\le\left|\cb_{\srk,\ell}\left(\bm{0},r\right)\right|\le K_{q}^{-\ell}\binom{\ell+r}{\ell}q^{\left(m+\eta-r/\ell\right)r}.
\end{equation*}
\end{propo}

Let us now provide the definition of \textit{sum-rank metric codes}, which can be classified into two types:\textit{General sum-rank metric codes} and \textit{linear sum-rank metric codes}.
\begin{de}[Sum-Rank Metric Code \cite{Mar2022}]
A \textit{general sum-rank metric code} is simply a non-empty subset of $\cm^{\ell}$. An $\mathbb{F}_q$-\textit{linear sum-rank metric code} forms an ($\mathbb{F}_q$-)subspace of $\cm^{\ell}$.
\end{de}

Let $\cc$ be a (general or $\mathbb{F}_q$-linear) sum-rank metric code. Then $(\cc,d_{\srk,\ell}\vert_{\cc\times\cc})$, where $d_{\srk,\ell}\vert_{\cc\times\cc}$ denotes the restriction of $d_{\srk,\ell}$ on $\cc\times\cc$, is a metric subspace of $(\cm^{\ell},d_{\srk,\ell})$. We shall denote $d_{\srk,\ell}\vert_{\cc\times\cc}$ simply as $d_{\srk,\ell}$ if there is no ambiguity. Let $\cc\le\cm^{\ell}$. The dimension of the code is then defined to be its dimension of it as a subspace of $\cm^{\ell}$.
The rate $R$ of a sum-rank metric code is $R:=\log_q\left(\left|\cc\right|\right)/mn$, where $n:=\eta\ell$ is the code length. For $\cc\le\cm^{\ell}$, the rate can be alternatively defined as $R:=\dim\left(\cc\right)/mn$ \cite{Che2021,Che2023}. 

There is already a great deal of excellent references comprehensively introducing sum-rank metric codes and their related properties, we recommend \cite{Byr2020,Ott2021,Mar2022,Ott2022,Ott2023,Gor2023}.

%Since this paper focuses on the list-decodability of random sum-rank metric codes, it is necessary to provide formal definitions of \textit{random sum-rank metric codes} and \textit{list-decodability}. Definitions of random Hamming metric codes and random rank metric codes have already appeared in related studies \cite{Res2019,Gur2017}. To the best of the authors' knowledge, there is \textit{no} existing literature that provides a formal definition for random sum-rank metric codes. We firmly believe that a compelling approach to defining random sum-rank metric codes is to follow the same principles used for defining random Hamming metric codes and random rank metric codes.

We now give formal definitions of \textit{random sum-rank metric codes} and their \textit{list-decodability}.
\begin{de}[Random Sum-Rank Metric Code]\label{de:1}
A \textit{random general sum-rank metric code} of rate $R$ is a uniformly random subset $\cc\subseteq\cm^{\ell}$ obtained by including each codeword independently with probability $q^{\left(R-1\right)mn}$. A \textit{random} $\mathbb{F}_q$-\textit{linear sum-rank metric code} of rate $R$ is defined to be a uniformly random subspace $\cc\le\cm^{\ell}$ of dimension $k:=Rmn$, which is assumed to be an integer.
\end{de}

Authors of \cite{Gur2006} provided an abstract definition of $\left(\rho,L\right)$-\textit{list-decodability}. Subsequently, authors of \cite{Din2014,Gur2017,Res2019} presented formal definitions of $\left(\rho, L\right)$-list-decodability for Hamming metric codes and rank metric codes, respectively. In \cite{Che2023}, the author provided the following definition for $\left(\rho, L\right)$-list-decodability of sum-rank metric codes.

\begin{de}[$\left(\rho, L\right)$-List-Decodability of Sum-Rank Metric Codes \cite{Che2023}]
Let $\cc$ be a (general or $\mathbb{F}_q$-linear) sum-rank metric code. For $\rho\in\left(0,1\right)$ and an integer $L\ge 1$, $\cc$ is said to be $\left(\rho,L\right)$-\textit{list-decodable}, if the number of codewords in $\cc$ within sum-rank distance $\rho n$ from any $\bm{X}\in\cm^{\ell}$ is at most $L$, that is:
\begin{equation*}
|\cb_{\srk,\ell}\left(\bm{X},\rho n\right)\cap\cc|\le L,\quad \forall\bm{X}\in\cm^{\ell}.
\end{equation*}
\end{de}

There is a lemma regarding the asymptotic order of binomial coefficients with a special form, that we give here since it will appear in the upper bounds of the volume of sum-rank metric balls.
\begin{lemma}\label{lem:8}
Let $a,b\in\mathbb{N}$ and $\alpha\in\left(0,1\right)$. Then for sufficiently large $a,b$, say $a,b=\Theta\left(k\right)$ for some sufficiently large $k$, the logarithm of binomial coefficients $\log_q(C^{\alpha ab+b}_{b})$ is at most of quasilinear order\footnote{Here and in what follows, to save space, we will sometimes use alternative notation for the binomial coefficient, $C_{\cdot}^{\cdot}$.}.
\end{lemma}

%Hi! alright, I will finish proofreading the document today and will probably be finished this evening. Then when you have made your changes (the gamma from 5. for example) I think we are ready to upload the first version.

%(I have copied what you have written above to my notes so its okay to remove it)

% Okay, I see, thanks so much!

\section{Decomposable Subspaces and Their Properties}\label{sec:2}
In this section, we will define the concept of the $\ell$-\textit{decomposable subspace}. We begin with its formal definition and then proceed to prove two fundamental properties associated with it.

Let $\cv_1,\cv_2,\ldots,\cv_{\ell}$ be $\eta$-dimensional $\mathbb{F}_q$-vector spaces. Let $\cu_{i}\le\cv_{i}$ be subspaces for each $i\in\left[1:\ell\right]$. It is easy to show that $\Pi_{i}^{\ell}\cu_i\le\Pi_{i}^{\ell}\cv_i$. Now we are interested in whether its converse is true, that is, if any subspace of $\Pi_{i}^{\ell}\cv_i$ can be written as a product of some subspaces $\cu_i\le\cv_i$ for all $i\in\left[1:\ell\right]$. Unfortunately, the converse is \textit{not} true in general, which is implied by Lemma \ref{lem:2}.
\begin{lemma}\label{lem:2}
Let $\eta$, $\ell$, and $w$ be positive integers. Then for all $q\ge 2$ we have:
\begin{equation}\label{eq:2}
\begin{bmatrix}
\eta\ell \\
w
\end{bmatrix}_q\ge\sum_{\bm{w}\in\ccp_{\ell}\left(w\right)}\prod_{i=1}^{\ell}\begin{bmatrix}
\eta \\ w_i
\end{bmatrix}_q
\end{equation}
\end{lemma}

Note that the left-hand side counts the number of all $w$-dimensional subspaces of the product space $\Pi_{i}^{\ell}\cv_i$, and the right-hand side counts the number of all $w$-dimensional subspaces which can be written as the product of $\cu_i\le\cv_i$ with $\dim\left(\cu_i\right)=w_i$ for all $i\in\left[1:\ell\right]$ such that $w_1+\ldots+w_{\ell}=w$, that is, the number of all $\ell$-\textit{decomposable} $w$-\textit{dimensional subspaces}.
\begin{de}[$\ell$-Decomposable Subspace]
Let $\cv_1,\cv_2,\ldots,\cv_{\ell}$ be $\eta$-dimensional $\mathbb{F}_q$-vector spaces. Let $\cu\le\Pi_{i}^{\ell}\cv_i$ be of dimension $w$. Then $\cu$ is said to be $\ell$-\textit{decomposable} if $\cu=\Pi_{i}^{\ell}\cu_i$, where $\cu_i\le\cv_i$ for all $i\in\left[1:\ell\right]$. The set of all $\ell$-\textit{decomposable} $w$-\textit{dimensional subspaces}, or simply $\ell$-\textit{decomposable subspaces} of $\Pi_{i}^{\ell}\cv_i$ will be denoted by $\cd_{\ell}(w,\Pi_{i}^{\ell}\cv_i)$.
\end{de}

We now present three fundamental properties of $\ell$-decomposable subspaces. These properties will facilitate the proofs of the \textit{dimension lemma} (cf. Lemma \ref{lem:7}) and Theorem \ref{th:4}.
\begin{lemma}[Basic Properties of $\ell$-Decomposable Subspaces]\label{lem:3}
Let $\cv$ be an $\mathbb{F}_q$-vector space of dimension $\eta$. Then:
\begin{enumerate}
\item $|\cd_{\ell}(w,\cv^{\ell})|$ is given by the right-hand side in \cref{eq:2}, and can be bounded from below and above by: 
\begin{equation*}
q^{\eta w-w^2/\ell}\le|\cd_{\ell}(w,\cv^{\ell})|\le K_q^{-\ell}\binom{w+\ell-1}{\ell-1}q^{\eta w-w^2/\ell}.
\end{equation*}
\item Let $\cx:=\Pi_{i}^{\ell}\cx_i$ and $\cy:=\Pi_{i}^{\ell}\cy_i$ be two $\ell$-decomposable spaces of $\cv^{\ell}$. Then the intersection $\cx\cap\cy$ is $\ell$-decomposable. Moreover:
\begin{equation*}
\dim\left(\cx\cap\cy\right)=\sum_{i=1}^{\ell}\dim\left(\cx_i\cap\cy_i\right).
\end{equation*}
\item Let $\cx\in\cd_{\ell}(w_x,\cv^{\ell})$ and $\cy\in\cd_{\ell}(w_y,\cv^{\ell})$, where $\cx:=\Pi_{i}^{\ell}\cx_i$ and $\cy:=\Pi_{i}^{\ell}\cy_i$. Then we have:
\begin{equation*}
\cx+\cy\in\cd_{\ell}(w_x+w_y-w_{xy},\cv^{\ell}),
\end{equation*}
where:
\begin{equation*}
w_{xy}:=\sum_{i=1}^{\ell}\dim\left(\cx_i\cap\cy_i\right).
\end{equation*}
That is, the sum of two $\ell$-decomposable subspaces is $\ell$-decomposable as well.
\end{enumerate}
\end{lemma}

\section{Main Results}\label{sec:3}
We have included all the main consequences of this paper in this section. In Theorem \ref{th:1} below we establish the first main result called the \textit{list-decoding capacity theorem for sum-rank metric codes}. Subsequently, we argue that the first part of this theorem is, in fact, equivalent to the \textit{list-decodability theorem for random general sum-rank metric codes}. Theorem \ref{th:1} is a generalization of the list-decoding capacity theorems for Hamming and rank metric codes. The proof idea and technique are consistent with that in \cite[Theorem III.1]{Din2014} and \cite[Theorem 2.4.7]{Res2019}.
\begin{theorem}[List-Decoding Capacity Theorem for Sum-Rank Metric Codes]\label{th:1}
Let $m$, $\eta$, and $\ell$ be positive integers satisfying $\eta\le m$\footnote{This assumption is without loss of generality, as we can consider the transpose code when $m\le \eta$.}. Define $b:=\eta/m$. Let $\rho\in\left(0,1\right)$ and $\epsilon>0$. Then:
\begin{enumerate}
\item There exists a sum-rank metric code $\cc$ in $\cm^{\ell}$ of rate $1-(\rho+\rho b-\rho^2b)-\epsilon$ which is $\left(\rho,O\left(1/\epsilon\right)\right)$-list-decodable.
\item For any sum-rank metric code $\cc$ in $\cm^{\ell}$ of rate $1-(\rho+\rho b-\rho^2b)+\epsilon$, there exists $\bm{X}\in\cm^{\ell}$ such that $|\cb_{\srk,\ell}(\bm{X},\rho n)\cap\cc|\ge q^{\Omega\left(mn\right)}$. Alternatively, any $\left(\rho,L\right)$-list-decodable sum-rank metric code of rate $1-(\rho+\rho b-\rho^2b)+\epsilon$ has $L\ge q^{\Omega\left(mn\right)}$.
\end{enumerate}
The quantity $1-\kappa_b\left(\rho\right):=1-(\rho+\rho b-\rho^2b)$ is said to be the \textit{list-decoding capacity of sum-rank metric codes}, and $\epsilon$ is called the \textit{capacity gap}.
\end{theorem}
\begin{pro}
Let $m$, $\eta$, and $\ell$ be positive integers satisfying $\eta\le m$. Define $b:=\eta/m$. Let $\rho\in\left(0,1\right)$ and $\epsilon>0$. 
\begin{enumerate}
\item Let $\cc\subseteq\cm^{\ell}$ be a random general sum-rank metric code of rate $R:=1-\left(\rho+\rho b-\rho^2b\right)-\epsilon$. We now bound the probability that $\cc$ fails $\left(\rho,L\right)$-list-decodability. If $\cc$ is \textit{not} $\left(\rho,L\right)$-list-decodable, then there is an $\bm{X}\in\cm^{\ell}$ and a subset $\ca\subseteq\cc$ with $\left|\ca\right|=L+1$ such that $\ca\subseteq\cb_{\srk,\ell}\left(\bm{X},\rho n\right)$. Then the probability that $\cc$ fails $\left(\rho,L\right)$-list-decodability is given by:

\begin{equation}\label{eq:41}
\pr[\exists\bm{X}\in\cm^{\ell},\ca\subseteq\cb_{\srk,\ell}\left(\bm{X},\rho n\right),\left|\ca\right|=L+1:\ca\subseteq\cc] \;
=
\sum_{\bm{X}\in\cm^{\ell}}\;\sum_{\substack{\ca\subseteq\cb_{\srk,\ell}\left(\bm{X},\rho n\right) \\ \left|\ca\right|=L+1}}\pr\left[\ca\subseteq\cc\right],
\end{equation}

where we take the union bound on all possible choices of $\bm{X}\in\cm^{\ell}$ and $\ca\subseteq\cb_{\srk,\ell}\left(\bm{X},\rho n\right)$ of size $\left|\ca\right|=L+1$. Note that by using the definition of random general sum-rank metric codes (cf. Definition \ref{de:1}) we have $\pr\left[\ca\subseteq\cc\right]=q^{\left(R-1\right)mn\left(L+1\right)}$. Thus, the probability in \cref{eq:41} equals and can be bounded from above by:

\begin{equation}\label{eq:42}
\begin{array}{ll}
\displaystyle{} & \displaystyle{\sum_{\bm{X}\in\cm^{\ell}}\;\sum_{\substack{\ca\subseteq\cb_{\srk,\ell}\left(\bm{X},\rho n\right) \\ \left|\ca\right|=L+1}}\pr\left[\ca\subseteq\cc\right]=|\cm^{\ell}|\binom{|\cb_{\srk,\ell}\left(\bm{0},\rho n\right)|}{L+1}q^{\left(R-1\right)mn\left(L+1\right)}} \vspace{1ex}\\
\displaystyle{\overset{\left(1\right)}{\le}} & \displaystyle{q^{mn}K_{q}^{-\ell\left(L+1\right)}\binom{\ell+\rho n}{\ell}^{L+1}q^{\left(m+\eta-\rho\eta\right)\rho n\left(L+1\right)}q^{\left(R-1\right)mn\left(L+1\right)}} \vspace{1ex}\\
\displaystyle{=} & \displaystyle{q^{mn}q^{-\ell\left(L+1\right)\log_q(K_q)+\left(L+1\right)\log_q(C_{\ell}^{\rho n+\ell})}q^{\left(m+\eta-\rho\eta\right)\rho n\left(L+1\right)}q^{\left(R-1\right)mn\left(L+1\right)}}\vspace{1ex}\\
\displaystyle{=} & \displaystyle{q^{mn}q^{((R+\rho+\rho b-\rho^2 b-1)mn-\ell\log_q\left(K_q\right)+\log_{q}(C^{\rho n+\ell}_{\ell}))\left(L+1\right)}} \vspace{1ex}\\ 
\displaystyle{\overset{\left(2\right)}{=}} & \displaystyle{q^{mn}q^{(-\epsilon mn-\ell\log_q\left(K_q\right)+\log_{q}(C^{\rho n+\ell}_{\ell}))\left(L+1\right)},}
\end{array}
\end{equation}

In inequality $\left(1\right)$ we use $|\cm^{\ell}|=(q^{m\eta})^{\ell}=q^{mn}$, $\binom{\beta}{\alpha} \le\beta^{\alpha}$, and the upper bound for the sum-rank metric ball given in Proposition \ref{prop:1}.
In equality $\left(2\right)$ we use the assumption $R=1-\left(\rho+\rho b-\rho^2b\right)-\epsilon$. 

If $L\ge 1/\epsilon$, say $L:=C/\epsilon$ for some $C\ge 1$, the probability in \cref{eq:42} is at most:
\begin{equation}\label{eq:8}
q^{\left(1-C-\epsilon\right)mn-(\ell\log_q\left(K_q\right)-\log_{q}(C^{\rho n+\ell}_{\ell}))\left(C/\epsilon+1\right)}.
\end{equation}
Define $T$ to be the negative exponent of $q$ in \cref{eq:8}, i.e.:
\begin{equation*}
T:=-\left(1-C-\epsilon\right)mn+(\ell\log_q\left(K_q\right)-\log_{q}(C^{\rho n+\ell}_{\ell}))\left(C/\epsilon+1\right).
\end{equation*} 
Note that, for sufficiently large\footnote{Here and subsequently, when we refer to sufficiently large $m$, $\eta$, and $\ell$, we mean that $m,\eta,\ell=\Theta\left(k\right)$ for some sufficiently large positive integer $k$. When we refer to $m$, $\eta$, and $\ell$ being sufficiently large compared to $1/\epsilon$, we mean that $m,\eta,\ell=\Theta\left(k\right)$ for some sufficiently large positive integer $k$ and $1/\epsilon \le m, \eta, \ell$.} $m$, $\eta$, and $\ell$ compared to $1/\epsilon$ the positive term $-\left(1-C-\epsilon\right)mn$ dominates and hence there exists some positive constant $C'$ such that $T\ge C'mn$, which implies that $T=\Omega\left(mn\right)$. Finally, $\cc$ is \textit{not} $\left(\rho,L\right)$-list-decodable with probability at most $q^{-T}=q^{-\Omega\left(mn\right)}$.
\item Let $\cc$ be a sum-rank metric code in $\cm^{\ell}$ of rate $R:=1-\left(\rho+\rho b-\rho^2b\right)+\epsilon$. Let $\bm{X}\in\cm^{\ell}$ be sampled uniformly at random. We compute the expectation:
\begin{equation}\label{eq:9}
\begin{array}{ll}
\displaystyle{\mathbb{E}\left[\left|\cb_{\srk,\ell}\left(\bm{X},\rho n\right)\cap\cc\right|\right]} & \displaystyle{=\sum_{\bm{C}\in\cc}\mathbb{E}\left[\rchi\left(\bm{C}\in\cb_{\srk,\ell}\left(\bm{X},\rho n\right)\right)\right]=\sum_{\bm{C}\in\cc}\mathbb{E}\left[\rchi\left(\bm{X}\in\cb_{\srk,\ell}\left(\bm{C},\rho n\right)\right)\right]} \vspace{1ex}\\
\displaystyle{} & \displaystyle{\overset{\left(1\right)}{=}\left|\cc\right||\cb_{\srk,\ell}\left(\bm{0},\rho n\right)|q^{-mn}}\\
\displaystyle{} & \displaystyle{\overset{\left(2\right)}{\ge} q^{Rmn}q^{(\rho+\rho b-\rho^2 b-1)mn+\ell\log_q\left(K_q\right)-\ell/4}}, \vspace{1ex}\\
\displaystyle{} & \displaystyle{\overset{\left(3\right)}{=}q^{\epsilon mn+\ell\log_q\left(K_q\right)-\ell/4},} 
\end{array}
\end{equation}
where $\rchi\left(\ce\right)$ represents the characteristic function of some event $\ce$. 
In equality $\left(1\right)$ we use the fact that $\mathbb{E}\left[\rchi\left(\ce\right)\right]=\bff{Pr}\left[\ce\right]$ and the uniformity of $\bm{X}\in\cm^{\ell}$. In inequality $\left(2\right)$ we use $\left|\cc\right|=q^{Rmn}$ and the lower bound for the sum-rank metric ball provided in Proposition \ref{prop:1}.
In equality $\left(3\right)$ we use the assumption $R=1-\left(\rho+\rho b-\rho^2b\right)+\epsilon$. For sufficiently large $m$, $\eta$, and $\ell$ we find that the positive term $\epsilon mn$ in the exponent dominates, and hence there is some positive constant $C''$ such that $\epsilon mn+\ell\log_q\left(K_q\right)-\ell/4\ge C''mn$, which means that the expectation in \cref{eq:9} is at least $q^{\Omega\left(mn\right)}$.
\hfill$\square$
\end{enumerate}
\end{pro}

In the proof of the first part in Theorem \ref{th:1}, the existence of the $\left(\rho,O\left(1/\epsilon\right)\right)$-list-decodable sum-rank metric code is guaranteed by constructing a random general sum-rank metric code. Thus, the first part in Theorem \ref{th:1} is indeed equivalent to Theorem \ref{th:2} below.
\begin{theorem}[List-Decodability for Random General Sum-Rank Metric Codes]\label{th:2}
Let $m$, $\eta$, and $\ell$ be positive integers satisfying $\eta\le m$. Define $b:=\eta/m$. Let $\rho\in\left(0,1\right)$ and $\epsilon>0$. Then any random general sum-rank metric code $\cc\subseteq\cm^{\ell}$ of rate $R=1-\kappa_b\left(\rho\right)-\epsilon$ is $\left(\rho,O\left(1/\epsilon\right)\right)$-list-decodable with high probability, that is:
\begin{equation*}
\pr\left[\cc\text{ is }\left(\rho,O\left(1/\epsilon\right)\right)\text{-list-decodable}\right]\ge 1-q^{-\Omega\left(mn\right)}.
\end{equation*}
\end{theorem}

By using the same proof technique as in the first part of Theorem \ref{th:1} we can derive a trivial version of the list-decodability theorem for random $\mathbb{F}_q$-linear sum-rank metric codes, as given by Theorem \ref{th:3}.
\begin{theorem}\label{th:3}
Let $m$, $\eta$, and $\ell$ be positive integers satisfying $\eta\le m$. Define $b:=\eta/m$. Let $\rho\in\left(0,1\right)$ and $\epsilon>0$. Then any random $\mathbb{F}_q$-linear sum-rank metric code $\cc\le\cm^{\ell}$ of rate $R=1-\kappa_b\left(\rho\right)-\epsilon$ is $\left(\rho,\exp\left(O\left(1/\epsilon\right)\right)\right)$-list-decodable with high probability, that is:
\begin{equation*}
\pr\left[\cc\,\,\text{is}\,\,\left(\rho,\exp\left(O\left(1/\epsilon\right)\right)\right)\text{-}\text{list\text{-}decodable}\right]\ge 1-q^{-\Omega\left(mn\right)}.
\end{equation*}
\end{theorem}
\begin{pro}
The codewords within $\cc$ now form a subspace of $\cm^\ell$ and are thus highly correlated. However, any subset of size $L+1$ within $\cc$ contains a subset $\ca$ containing at least $\log_q\left(L+1\right)$ linearly independent vectors. Then, by using the same proof technique as in the first part of Theorem \ref{th:1} we note that the probability that $\cc$ fails the $\left(\rho,L\right)$-list-decodability is at most $q^{-\Omega\left(mn\right)}$ if $\log_q\left(L+1\right)=C/\epsilon$ for some $C\ge 1$, that is, if $L+1=q^{C/\epsilon}$, or equivalently, $L=\exp\left(O\left(1/\epsilon\right)\right)$.
\hfill$\square$
\end{pro}

In Theorem \ref{th:3}, we clearly observe that the list size for random linear sum-rank metric codes is currently exponentially large. As mentioned in the proof, this arises from the high correlation among codewords in the linear code, specifically due to the closedness of $\cc$ with respect to vector space addition, that is, for $\bm{C}_1,\bm{C}_2\in\cc$ we have $\bm{C}_1+\bm{C}_2\in\cc$. Therefore, we need to establish an auxiliary result to help ``mitigate" this correlation between codewords. We refer to this result as the \textit{limited correlation property between sum-rank metric balls and subspaces} (cf. Lemma \ref{lem:6}), which states that if $\gamma$ elements are sampled uniformly and independently at random from $\cb_{\srk,\ell}\left(\bm{0},\rho n\right)$, then the probability that $\Omega\left(\gamma\right)$ elements in their linear span, which forms a subspace, lie in the same ball is exponentially small. Lemma \ref{lem:6} generalizes the corresponding results for Hamming metric codes \cite[Theorem 3]{Gur2010} and rank metric codes \cite[Lemma 4.2]{Gur2017}. To prove this limited correlation property, we first need to establish some auxiliary results.

First, we would like to show the dimension lemma, which states that if $\cx$ and $\cy$ are independently and uniformly random $\ell$-decomposable subspaces of $\mathbb{F}_q^{\eta\ell}$ of dimensions $w_x$ and $w_y$, respectively, the probability that the dimension of their intersection is greater than zero is exponentially small for sufficiently large $\eta$ and $\ell$. The proof idea is as follows: First we outline a sampling method that produces a distribution, which is equivalent to the uniform distribution over $\cd_{\ell}(w,\mathbb{F}_q^{\eta\ell})$. Then we will show that $\mathbb{E}\left[\dim\left(\cx\cap\cy\right)\right]$ is exponentially small. Finally, the exponentially small expectation implies the exponentially small probability, since $\dim\left(\cx\cap\cy\right)$ is a discrete random variable defined on some finite subset of $\mathbb{N}_0$.
\begin{theorem}\label{th:a1}
Let $\eta$, $\ell$, $w_x$, $w_y$ be positive integers and $w_x,w_y=\Theta\left(\eta\ell\right)$. Let $\cx$ and $\cy$ be independently and uniformly random $\ell$-decomposable subspaces of $\mathbb{F}_q^{\eta\ell}$ of dimensions $w_x$ and $w_y$, respectively. Then:
\begin{equation*}
\mathbb{E}\left[\dim\left(\cx\cap\cy\right)\right]\le q^{-\Omega\left(\eta n\right)}.
\end{equation*}
\end{theorem}
\begin{pro}
Sampling $\cu:=\Pi_{i}^{\ell}\cu_i\in\cd_{\ell}(w,\mathbb{F}^{\eta\ell}_q)$ uniformly at random means that sample every element with equal probability $1/|\cd_{\ell}(w,\mathbb{F}^{\eta\ell}_q)|$. For each $\bm{w}\in\ccp_{\ell}\left(w\right)$, define $\cd_{\ell}(w,\mathbb{F}^{\eta\ell}_q,\bm{w})$ be the set of all $\ell$-decomposable subspaces associated with the partition $\bm{w}$. Clearly, $\{\cd_{\ell}(w,\mathbb{F}^{\eta\ell}_q,\bm{w})\mid\bm{w}\in\ccp_{\ell}\left(w\right)\}$ forms a partition of $\cd_{\ell}(w,\mathbb{F}^{\eta\ell}_q)$. Now, we would like to design a new sampling procedure, which describes a distribution which is equivalent to uniform distribution over $\cd_{\ell}(w,\mathbb{F}_q^{\eta\ell})$. Let $\bm{W}$ be the random variable specifying the partition and $U$ be the random variable specifying the decomposable subspace (whose value depends on the choice of $\bm{W}$). Consider the following procedure:
\begin{enumerate}
\item Sample $\bm{w}\in\ccp_{\ell}\left(w\right)$ with probability $\pr\left[\bm{W}=\bm{w}\right]=\prod_{i=1}^{\ell}\stirling{\eta}{w_i}_q/|\cd_{\ell}(w,\mathbb{F}^{\eta\ell}_q)|$.
\item Conditioning on $\left\{\bm{W}=\bm{w}\right\}$, sample $\cu:=\Pi_{i}^{\ell}\cu_i\in\cd_{\ell}(w,\mathbb{F}^{\eta\ell}_q,\bm{w})$ uniformly at random, that is sample $\cu$ from $\cd_{\ell}(w,\mathbb{F}^{\eta\ell}_q,\bm{w})$ with probability $\pr\left[U=\cu\mid\bm{W}=\bm{w}\right]=(\prod_{i=1}^{\ell}\stirling{\eta}{w_i}_q)^{-1}$.
\end{enumerate}

Clearly, this sampling procedure gives a pair $\left(\cu,\bm{w}\right)$ with probability $\pr\left[U=\cu,\bm{W}=\bm{w}\right]=1/|\cd_{\ell}(w,\mathbb{F}^{\eta\ell}_q)|$, which corresponds to a unique $\ell$-decomposable subspace in $\cd_{\ell}(w,\mathbb{F}_q^{\eta\ell})$. Actually, one can identify $\left(\cu,\bm{w}\right)$ as $\cu\in\cd_{\ell}(w,\mathbb{F}_q^{\eta\ell})$. Thus, the distribution delineated through the aforementioned sampling process is equivalent to the uniform distribution over $\cd_{\ell}(w,\mathbb{F}_q^{\eta\ell})$. We also note that uniform $\cu\in\cd_{\ell}(w,\mathbb{F}^{\eta\ell}_q,\bm{w})$ is equivalent to independently uniform $\cu_i\in\cg_{\mathbb{F}_q}\left(\eta,w_i\right)$ for all $i\in\left[1:\ell\right]$.

Let $X$ and $Y$ be independently uniform random variables over $\cd_{\ell}(w_x,\mathbb{F}_q^{\eta\ell})$ and $\cd_{\ell}(w_y,\mathbb{F}_q^{\eta\ell})$, respectively. Then:
\begin{equation*}
\begin{array}{ll}
\displaystyle{} & \displaystyle{\mathbb{E}\left[\dim\left(X\cap Y\right)\right]=\sum_{\substack{\cx\in\cd_{\ell}(w_x,\mathbb{F}_q^{\eta\ell}) \\ \cy\in\cd_{\ell}(w_y,\mathbb{F}_q^{\eta\ell})}}\dim\left(\cx\cap\cy\right)\pr\left[X=\cx,Y=\cy\right]} \vspace{1ex}\\
\displaystyle{=} & \displaystyle{\sum_{d\in\left[0:\min\{w_x,w_y\}\right]}d\cdot\sum_{\substack{
\cx\in\cd_{\ell}(w_x,\mathbb{F}_q^{\eta\ell})\\ \cy\in\cd_{\ell}(w_y,\mathbb{F}_q^{\eta\ell})\\\dim\left(\cx\cap\cy\right)=d}}
\pr\left[X=\cx,Y=\cy\right].}
\end{array}
\end{equation*}
Since $\{\cd_{\ell}(w,\mathbb{F}^{\eta\ell}_q,\bm{w})\mid\bm{w}\in\ccp_{\ell}\left(w\right)\}$ forms a partition of $\cd_{\ell}(w,\mathbb{F}^{\eta\ell}_q)$, and $\pr\left[U=\cu\right]=\pr\left[S=\cu,\bm{W}=\bm{w}\right]$\footnote{Note that $U$ is a uniform random variable defined on $\cd_{\ell}(w,\mathbb{F}_q^{\eta\ell})$ and $S:=S_{\bm{W}=\bm{w}}$ is a uniform random variable defined on $\cd_{\ell}(w,\mathbb{F}_q^{\eta\ell},\bm{w})$.} if $\cu$ is uniformly distributed on $\cd_{\ell}(w,\mathbb{F}_q^{\eta\ell})$ and $\left(\cu,\bm{w}\right)$ follows the distribution described by the sampling process mentioned above, then:
\begin{equation*}
 \sum_{\substack{
\cx\in\cd_{\ell}(w_x,\mathbb{F}_q^{\eta\ell})\\ \cy\in\cd_{\ell}(w_y,\mathbb{F}_q^{\eta\ell})\\\dim\left(\cx\cap\cy\right)=d}}
\pr\left[X=\cx,Y=\cy\right]\;
=
\sum_{\substack{\bm{\mu}\in\ccp_{\ell}\left(w_x\right)\\ \bm{\nu}\in\ccp_{\ell}\left(w_y\right)}}\,\sum_{
\substack{\cx\in\cd_{\ell}(w_x,\mathbb{F}_q^{\eta\ell},\bm{\mu})\\\cy\in\cd_{\ell}(w_y,\mathbb{F}_q^{\eta\ell},\bm{\nu})
\\\dim\left(\cx\cap\cy\right)=d}
}\pr\left[\left(S,\bm{W}_x\right)=\left(\cx,\bm{\mu}\right),\left(T,\bm{W}_y\right)=\left(\cy,\bm{\nu}\right)\right],
\end{equation*}
where $S:=S_{\bm{\mu}}$ and $T:=T_{\bm{\nu}}$ are uniform random variables defined on $\cd_{\ell}(w_x,\mathbb{F}_q^{\eta\ell},\bm{\mu})$ and $\cd_{\ell}(w_y,\mathbb{F}_q^{\eta\ell},\bm{\nu})$, respectively. Moreover, we note that the event $\left\{\left(S,\bm{W}_x\right)=\left(\cx,\bm{\mu}\right),\left(T,\bm{W}_y\right)=\left(\cy,\bm{\nu}\right)\right\}$ is equivalent to the event $\left\{S=\cx,\bm{W}_x=\bm{\mu},T=\cy,\bm{W}_y=\bm{\nu}\right\}$, then:
\begin{equation*}
\begin{array}{ll}
& \displaystyle{\sum_{
\substack{\cx\in\cd_{\ell}(w_x,\mathbb{F}_q^{\eta\ell},\bm{\mu})\\\cy\in\cd_{\ell}(w_y,\mathbb{F}_q^{\eta\ell},\bm{\nu})
\\\dim\left(\cx\cap\cy\right)=d}
}\pr\left[\left(S,\bm{W}_x\right)=\left(\cx,\bm{\mu}\right),\left(T,\bm{W}_y\right)=\left(\cy,\bm{\nu}\right)\right]} \vspace{1ex}\\
\displaystyle{=} & \displaystyle{\sum_{
\substack{\cx\in\cd_{\ell}(w_x,\mathbb{F}_q^{\eta\ell},\bm{\mu})\\\cy\in\cd_{\ell}(w_y,\mathbb{F}_q^{\eta\ell},\bm{\nu})
\\\dim\left(\cx\cap\cy\right)=d}
}\pr\left[S=\cx,T=\cy\mid\bm{W}_x=\bm{\mu},\bm{W}_y=\bm{\nu}\right]\pr\left[\bm{W}_x=\bm{\mu},\bm{W}_y=\bm{\nu}\right]} \vspace{1ex}\\
\displaystyle{=:} & \displaystyle{\pr\left[\dim\left(S\cap T\right)=d\mid \bm{W}_x=\bm{\mu},\bm{W}_y=\bm{\nu}\right]\pr\left[\bm{W}_x=\bm{\mu},\bm{W}_y=\bm{\nu}\right].}
\end{array}
\end{equation*}
Thus $\mathbb{E}\left[\dim\left(X\cap Y\right)\right]$ equals:
\begin{equation*}
\begin{array}{ll}
 & \displaystyle{\sum_{\substack{\bm{\mu}\in\ccp_{\ell}\left(w_x\right)\\\bm{\nu}\in\ccp_{\ell}\left(w_y\right)}}\left(\sum_{d\in\left[0:\min\{w_x,w_y\}\right]}d\pr\left[\dim\left(S\cap T\right)=d\mid \bm{W}_x=\bm{\mu},\bm{W}_y=\bm{\nu}\right]\right)\pr\left[\bm{W}_x=\bm{\mu},\bm{W}_y=\bm{\nu}\right]} \vspace{1ex}\\
\displaystyle{=:} & \displaystyle{\sum_{\substack{\bm{\mu}\in\ccp_{\ell}\left(w_x\right)\\\bm{\nu}\in\ccp_{\ell}\left(w_y\right)}}\mathbb{E}\left[\dim\left(S\cap T\right)\mid\bm{W}_x=\bm{\mu},\bm{W}_y=\bm{\nu}\right]\pr\left[\bm{W}_x=\bm{\mu},\bm{W}_y=\bm{\nu}\right]}
\end{array}
\end{equation*}
Using Lemma \ref{lem:3}.2 and the linearity of the conditional expectation gives:
\begin{equation*}
\begin{array}{ll}
\displaystyle{} & \displaystyle{\mathbb{E}\left[\dim\left(S\cap T\right)\mid\bm{W}_x=\bm{\mu},\bm{W}_y=\bm{\nu}\right]=\sum_{i\in\left[1:\ell\right]}\mathbb{E}\left[\dim\left(S_i\cap T_i\right)\mid\bm{W}_x=\bm{\mu},\bm{W}_y=\bm{\nu}\right]} \vspace{1ex}\\
\displaystyle{=} & \displaystyle{\sum_{i\in\left[1:\ell\right]}\left(\sum_{d_i\in\left[0:\min\{w_{x,i},w_{y,i}\}\right]}d_i\pr\left[\dim\left(S_i\cap T_i\right)=d_i\mid\bm{W}_x=\bm{\mu},\bm{W}_y=\bm{\nu}\right]\right).}
\end{array}
\end{equation*}
Now we would like to provide an upper bound for $\mathbb{E}\left[\dim\left(S\cap T\right)\mid\bm{W}_x=\bm{\mu},\bm{W}_y=\bm{\nu}\right]$. Conditioning on the event $\left\{\bm{W}_x=\bm{\mu},\bm{W}_y=\bm{\nu}\right\}$ note that $S_i$ and $T_i$ are independently uniform random variables over $\cg_{\mathbb{F}_q}\left(\eta,w_{x,i}\right)$ and $\cg_{\mathbb{F}_q}\left(\eta,w_{y,i}\right)$ respectively for each $i\in\left[1:\ell\right]$. For each $i\in\left[1:\ell\right]$ we fix $S_i$, i.e., set $S_i=\cx_i\in\cg_{\mathbb{F}_q}\left(\eta,w_{x,i}\right)$. Clearly, the event $\{\exists\cu_i\in\cg_{\mathbb{F}_q}\left(d_i,\cx_i\right):\cu_i\le T_i\mid\bm{W}_x=\bm{\mu},\bm{W}_y=\bm{\nu}\}$ is equivalent to the event $\{\dim\left(S_i\cap T_i\right)\ge d_i\mid\bm{W}_x=\bm{\mu},\bm{W}_y=\bm{\nu}\}$ and hence contains the event $\{\dim\left(S_i\cap T_i\right)=d_i\mid\bm{W}_x=\bm{\mu},\bm{W}_y=\bm{\nu}\}$ for each $i\in\left[1:\ell\right]$ and each $d_i\in\left[0:\min\{w_{x,i},w_{y,i}\}\right]$\footnote{The proof idea of this part is inspired by \cite[Claim 1]{Gur2017}.}. Thus, by the monotonicity of thr probability measure we find that:
\begin{equation}\label{eq:a1}
\begin{array}{ll}
\displaystyle{} & \displaystyle{\sum_{i\in\left[1:\ell\right]}\sum_{d_i\in\left[0:\min\{w_{x,i},w_{y,i}\}\right]}d_i\pr\left[\dim\left(S_i\cap T_i\right)=d_i\mid\bm{W}_x=\bm{\mu},\bm{W}_y=\bm{\nu}\right]} \vspace{1ex}\\
\displaystyle{\le} & \displaystyle{\sum_{i\in\left[1:\ell\right]}\sum_{d_i\in\left[0:\min\{w_{x,i},w_{y,i}\}\right]}d_i\pr\left[\exists\cu_i\in\cg_{\mathbb{F}_q}\left(d_i,\cx_i\right):\cu_i\le T_i\mid\bm{W}_x=\bm{\mu},\bm{W}_y=\bm{\nu}\right]} \vspace{1ex}\\
\displaystyle{\overset{\left(1\right)}{\le}} & \displaystyle{\sum_{i\in\left[1:\ell\right]}\sum_{d_i\in\left[0:\min\left\{w_{x,i},w_{y,i}\right\}\right]}d_i\sum_{\cu_i\in\cg_{\mathbb{F}_q}\left(d_i,\cx_i\right)}\pr\left[\cu_i\le T_i\mid\bm{W}_x=\bm{\mu},\bm{W}_y=\bm{\nu}\right]} \vspace{1ex}\\
\displaystyle{\overset{\left(2\right)}{=}} & \displaystyle{\sum_{i\in\left[1:\ell\right]}\sum_{d_i\in\left[0:\min\left\{w_{x,i},w_{y,i}\right\}\right]}d_i\left.\begin{bmatrix}
w_{x,i} \\ d_i 
\end{bmatrix}_q\begin{bmatrix}
\eta-d_i \\ w_{y,i}-d_i
\end{bmatrix}_q\right/\begin{bmatrix}
\eta \\ w_{y,i}
\end{bmatrix}_q} \vspace{1ex}\\
\displaystyle{\overset{\left(3\right)}{\le}} & \displaystyle{\sum_{i\in\left[1:\ell\right]}\sum_{d_i\in\left[0:\min\left\{w_{x,i},w_{y,i}\right\}\right]}d_iK_q^{-2}q^{-d_i(\eta-w_{y,i}-w_{x,i}+d_i)},}
\end{array}
\end{equation}
where in inequality (1) we use union bound, in equality (2) we use the uniformity of $T_i$ conditioning on $\{\bm{W}_y=\bm{\nu}\}$ and Lemma \ref{lem:1}.2, and in inequality (3) in we use upper and lower bounds of $q$-binomial coefficients in Lemma \ref{lem:1}.1. We also note that by independence of $\{\bm{W}_x=\bm{\mu}\}$ and $\{\bm{W}_y=\bm{\nu}\}$:

\begin{align*}
\pr\left[\bm{W}_x=\bm{\mu},\bm{W}_y=\bm{\nu}\right]
&=
\frac{\prod_{i=1}^{\ell}
\begin{bmatrix}
\eta\\w_{x,i}
\end{bmatrix}_q\prod_{i=1}^{\ell}
\begin{bmatrix}
\eta\\w_{y,i}
\end{bmatrix}_q}{|\cd_{\ell}(w_x,\mathbb{F}^{\eta\ell}_q)||\cd_{\ell}(w_y,\mathbb{F}^{\eta\ell}_q)|}\\[2ex]
&\overset{\left(1\right)}{\le} 
\frac{K_q^{-2\ell}q^{\eta w_x-\xi_x+\eta w_y-\xi_y}}{q^{\eta w_x-w^{2}_x/\ell+\eta w_y-w^2_y/\ell}}
=
K_q^{-2\ell}q^{-(\xi_x+\xi_y-w^2_y/\ell-w^{2}_x/\ell)},
\end{align*}

where in inequality (1) we define $\xi_j:=\sum_{i\in\left[1:\ell\right]}w_{j,i}^2$ for $j\in\{x,y\}$ and we use $\sum_{i\in\left[1:\ell\right]}w_{j,i}=w_j$, the lower bound of $|\cd_{\ell}(w_{j},\mathbb{F}^{\eta\ell}_q)|$ provided in Lemma \ref{lem:3}.1 and the upper bound of $\stirling{\eta}{w_j}_q$ in Lemma \ref{lem:1}.1. Note that by the Cauchy-Schwarz inequality we have:
\begin{equation*}
\xi_{j}=\left(\sum_{i\in\left[1:\ell\right]}w_{j,i}^2\right)1=\left(\sum_{i\in\left[1:\ell\right]}w_{j,i}^2\right)\left(\sum_{i\in\left[1:\ell\right]}\frac{1}{\ell}\right)\ge\left(\sum_{i\in\left[1:\ell\right]}\frac{w_{j,i}}{\sqrt{\ell}}\right)^2=\frac{w_{j}^2}{\ell}.
\end{equation*}
Thus, for $w_{j}=\Theta\left(\eta\ell\right)$, $\xi_{j}-w_{j}^2/\ell\ge \Omega(\eta^2\ell)$ for $j\in\left\{x,y\right\}$. Finally,
\begin{align*}
    \mathbb{E}\left[\dim\left(X\cap Y\right)\right] & \leq
    \sum_{\substack{\bm{\mu}\in\ccp_{\ell}\left(w_x\right)\\\bm{\nu}\in\ccp_{\ell}\left(w_y\right)}}\,\sum_{i\in\left[1:\ell\right]}\;\sum_{d_i\in\left[0:\min\left\{w_{x,i},w_{y,i}\right\}\right]}q^{-\Omega(\eta^2\ell)-d_i(\eta-w_{y,i}-w_{x,i}+d_i)+\log_q(d_i)-2\left(\ell+1\right)\log_q(K_q)}\\[1.0ex]
    & \overset{\left(1\right)}{\le}\ell\binom{w_x+\ell-1}{\ell-1}\binom{w_y+\ell-1}{\ell-1}\left(\eta+1\right)q^{-\Omega(\eta^2\ell)}=q^{-\Omega(\eta^2\ell)},
\end{align*}

where in inequality (1) we use the fact that $-d_i(\eta-w_{y,i}-w_{x,i}+d_i)+\log_q(d_i)-2\left(\ell+1\right)\log_q(K_q)$ is at most of quadratic order in $\eta$ and $\ell$. 
\hfill$\square$
\end{pro}

A direct consequence of Theorem \ref{th:a1} is the following:
\begin{cor}\label{cor:a1}
Let $\eta$, $\ell$, $w_x$, $w_y$ be positive integers and $w_x,w_y=\Theta\left(\eta\ell\right)$. Let $\cx$ and $\cy$ be independently and uniformly random $\ell$-decomposable subspaces of $\mathbb{F}_q^{\eta\ell}$ of dimensions $w_x$ and $w_y$, respectively. Then for all $d\in\left[1:\min\left\{w_x,w_y\right\}\right]$ we have:
\begin{equation*}
\pr\left[\dim\left(\cx\cap\cy\right)=d\right]\le q^{-\Omega\left(\eta n\right)}.
\end{equation*}
\end{cor}
\begin{pro}
Since $\dim\left(\cx\cap\cy\right)$ is a random variable defined on $\left[0:\min\left\{w_x,w_y\right\}\right]$, we have:

\begin{align*}
    q^{-\Omega(\eta^2\ell)}\ge\mathbb{E}\left[\dim\left(\cx\cap\cy\right)\right] & =\sum_{d\in\left[1:\min\left\{w_x,w_y\right\}\right]}d\pr\left[\dim\left(\cx\cap\cy\right)=d\right]\\[1.0ex]
    & \geq \sum_{d\in\left[1:\min\left\{w_x,w_y\right\}\right]}\pr\left[\dim\left(\cx\cap\cy\right)=d\right].
\end{align*}

\noindent Suppose that there is some $d_0\in\left[1:\min\left\{w_x,w_y\right\}\right]$ such that $\pr\left[\dim\left(\cx\cap\cy\right)=d_0\right]>q^{-\Omega(\eta^2\ell)}$. Then:
\begin{equation*}
q^{-\Omega(\eta^2\ell)}\ge\sum_{d\in\left[1:\min\left\{w_x,w_y\right\}\right]}\pr\left[\dim\left(\cx\cap\cy\right)=d\right]\ge\pr\left[\dim\left(\cx\cap\cy\right)=d_0\right]>q^{-\Omega(\eta^2\ell)}
\end{equation*}
which is a contradiction.
\hfill$\square$
\end{pro}

Finally, we can establish the dimension lemma using Corollary \ref{cor:a1}.

\begin{lemma}[Dimension Lemma]\label{lem:7}
Let $\eta$, $\ell$, $w_x$, $w_y$ be positive integers and $w_x,w_y=\Theta\left(\eta\ell\right)$. Suppose without loss of generality that $w_x\le w_y$. Let $\cx$ and $\cy$ be independently and uniformly random $\ell$-decomposable subspaces of $\mathbb{F}_q^{\eta\ell}$ of dimensions $w_x$ and $w_y$, respectively. Then for all $\alpha\in\left(0,1\right)$ we have:
\begin{equation*}
\pr\left[\dim\left(\cx\cap\cy\right)\ge\alpha w_x\right]\le q^{-\Omega\left(\eta n\right)}.
\end{equation*}
\end{lemma}
\begin{pro}
Note that the event $\{\dim\left(\cx\cap\cy\right)\ge\alpha w_x\}$ is equivalent to the event $\{\exists d\in\left[\alpha w_x:w_x\right]:\dim\left(\cx\cap\cy\right)=d\}$. Then we have:

\begin{align*}
    \pr\left[\dim\left(\cx\cap\cy\right)\ge\alpha w_x\right] &=\pr\left[\exists d\in\left[\alpha w_x:w_x\right]:\dim\left(\cx\cap\cy\right)=d\right]\\[1.0ex]
    & \leq \sum_{d\in\left[\alpha w_x:w_x\right]}\pr\left[\dim\left(\cx\cap\cy\right)=d\right]\le\left(1-\alpha\right)w_x q^{-\Omega(\eta^2\ell)}=q^{-\Omega(\eta^2\ell)}
\end{align*}

for sufficiently large $\eta$ and $\ell$.
\hfill$\square$
\end{pro}

Using the dimension lemma, we can prove the following core theorem, which states that for $\bm{X}_1$ and $\bm{X}_2$ sampled independently and uniformly at random from $\cb_{\srk,\ell}(\bm{0},\rho n)$, the probability that their sum falls in any ball $\cb_{\srk,\ell}(\bm{Y},\rho n)$, where $\bm{Y}\in\cm^{\ell}$, is exponentially small. Theorem \ref{th:4} generalizes the corresponding results of Hamming metric codes \cite[Lemma 7]{Gur2010} and rank metric codes \cite[Lemma 4.1]{Gur2017}. The proof technique is consistent with that in \cite[Lemma 4.1]{Gur2017} and \cite[Lemma 5.5.4]{Res2019}.
\begin{theorem}\label{th:4}
Let $m$, $\eta$, and $\ell$ be positive integers. Suppose that $\eta\le m$ and define $b:=\eta/m$. Let $\bm{Y}\in\cm^{\ell}$ be fixed. Let $\rho\in\left(0,1\right)$. Let $\bm{X}_1,\bm{X}_2\sim\mathfrak{D}_1$ denote the distribution where $\bm{X}_1$ and $\bm{X}_2$ are sampled independently and uniformly at random from $\cb_{\srk,\ell}\left(\bm{0},\rho n\right)$. Then for sufficiently large $m$, $\eta$, and $\ell$ we have:
\begin{equation*}
\pr_{\bm{X}_1,\bm{X}_2\sim\mathfrak{D}_1}\left[\bm{X}_1+\bm{X}_2\in\cb_{\srk,\ell}\left(\bm{Y},\rho n\right)\right]\le q^{-\Omega\left(mn\right)}.
\end{equation*}
\end{theorem}
\begin{pro}
Without loss of generality we assume that $\rho n$ is an integer. Denote the probability of interest as $p$. Let $\left(w_1,w_2\right)\in\left[0:\rho n\right]\times\left[0:\rho n\right]$ be the pair of sum-rank weights maximizing the probability below:
\begin{equation*}
\pr_{\bm{X}_1,\bm{X}_2\sim\mathfrak{D}_1}\left[\left(\bm{X}_1+\bm{X}_2\in\cb_{\srk,\ell}\left(\bm{Y},\rho n\right)\right)\cap\ce\left(t_1,t_2\right)\right],
\end{equation*}
where $\ce\left(t_1,t_2\right)$ denotes the event that $\wt_{\srk,\ell}\left(\bm{X}_1\right)=t_1$ and $\wt_{\srk,\ell}\left(\bm{X}_2\right)=t_2$. Thus:

\begin{equation}\label{eq:14}
\begin{array}{ll}
\displaystyle{p} & \displaystyle{\le\sum_{\left(t_1,t_2\right):\ce\left(t_1,t_2\right)}\pr_{\bm{X}_1,\bm{X}_2\sim\mathfrak{D}_1}\left[\left(\bm{X}_1+\bm{X}_2\in\cb_{\srk,\ell}\left(\bm{Y},\rho n\right)\right)\cap\ce\left(w_1,w_2\right)\right]} \vspace{1ex}\\
\displaystyle{} & \displaystyle{\overset{\left(1\right)}{\le}n^2\pr_{\bm{X}_1,\bm{X}_2\sim\mathfrak{D}_1}\left[\left(\bm{X}_1+\bm{X}_2\in\cb_{\srk,\ell}\left(\bm{Y},\rho n\right)\right)\cap\ce\left(w_1,w_2\right)\right]},
\end{array}
\end{equation}
where the inequality $\left(1\right)$ holds since there are at most $n^2$ ordered pairs $\left(t_1,t_2\right)\in\left[0:\rho n\right]\times\left[0:\rho n\right]$. A direct consequence of \cref{eq:14} is:
\begin{equation}\label{eq:15}
p\le n^2\pr_{\bm{X}_1,\bm{X}_2\sim\mathfrak{D}_1}\left[\bm{X}_1+\bm{X}_2\in\cb_{\srk,\ell}\left(\bm{Y},\rho n\right)\vert \ce\left(w_1,w_2\right)\right].
\end{equation}
Now, let us consider the following two cases: (1): $0\le w_1,w_2\le\left(1-\delta\right)\rho n$, and (2): $\left(1-\delta\right)\rho n\le w_1,w_2\le\rho n$ for some $\delta\in\left(0,1\right)$. For the first case note from \cref{eq:14} that:
\begin{equation}\label{eq:37}
\begin{array}{ll}
\displaystyle{p} & \displaystyle{\le n^2\pr_{\bm{X}_1,\bm{X}_2\sim\mathfrak{D}_1}\left[\left(\wt_{\srk,\ell}\left(\bm{X}_1\right)\le\left(1-\delta\right)\rho n\right)\cup\left(\wt_{\srk,\ell}\left(\bm{X}_2\right)\le\left(1-\delta\right)\rho n\right)\right]} \vspace{1ex}\\
\displaystyle{} & \displaystyle{\le n^2\sum_{j=1}^{2}\pr_{\bm{X}_1,\bm{X}_2\sim\mathfrak{D}_1}\left[\wt_{\srk,\ell}\left(\bm{X}_j\right)\le\left(1-\delta\right)\rho n\right].}
\end{array}
\end{equation}
Since $\wt_{\srk,\ell}\left(\bm{X}_j\right)\le\left(1-\delta\right)\rho n$ is equivalent to $\bm{X}_j\in\cb_{\srk,\ell}\left(\bm{0},\left(1-\delta\right)\rho n\right)$ for each $j\in\left[1:2\right]$, then by uniformity we obtain:
\begin{equation}\label{eq:16}
\pr_{\bm{X}_1,\bm{X}_2\sim\mathfrak{D}_1}\left[\wt_{\srk,\ell}\left(\bm{X}_j\right)\le\left(1-\delta\right)\rho n\right]=\frac{\left|\cb_{\srk,\ell}\left(\bm{0},\left(1-\delta\right)\rho n\right)\right|}{\left|\cb_{\srk,\ell}\left(\bm{0},\rho n\right)\right|}
\end{equation}
for each $j\in\left[1:2\right]$. By using the upper bound of numerator and the lower bound of denominator (cf. Proposition \ref{prop:1}) the ratio of the volumes of the two balls in \cref{eq:16} has the following upper bound:
\begin{equation}\label{eq:17}
q^{-mn(\delta\rho+\delta\rho b+\delta^2\rho^2b-2\delta\rho^2b)+\ell/4+\log_q(C^{\ell+\left(1-\delta\right)\rho n}_{\ell})-2\ell\log_{q}\left(K_q\right)}.
\end{equation}
Define $T_1$ to be the negative exponent of $q$ in \cref{eq:17}. Then we would like to show that $T_1=\Omega\left(mn\right)$. Note that:
\begin{equation*}
\delta\rho+\delta\rho b+\delta^2\rho^2b-2\delta\rho^2b=\delta\rho\left(1-\rho b\right)+\delta\rho b\left(1-\rho\right)+\delta^2\rho^2b>0,
\end{equation*}
since $\delta,\rho\in\left(0,1\right)$ and $b\in\left(0,1\right]$. Thus, the positive term $mn\left(\delta\rho+\delta\rho b+\delta^2\rho^2b-2\delta\rho^2b\right)$ dominates for sufficiently large $m$, $\eta$, and $\ell$. Thus, one can find a positive constant $C$ such that $T_1\ge Cmn$, which implies that $T_1=\Omega\left(mn\right)$. Finally, by \cref{eq:17} we obtain:
\begin{equation*}
\frac{\left|\cb_{\srk,\ell}\left(\bm{0},\left(1-\delta\right)\rho n\right)\right|}{\left|\cb_{\srk,\ell}\left(\bm{0},\rho n\right)\right|}\le q^{-T_1}=q^{-\Omega\left(mn\right)}.
\end{equation*}
Then by \cref{eq:37} and \cref{eq:16} we obtain:
\begin{equation*}
p\le 2n^2q^{-\Omega\left(mn\right)}=\left(1/q\right)^{(\Omega\left(mn\right)-2\log_q(\sqrt{2}n))}=q^{-\Omega\left(mn\right)}.
\end{equation*}
Now let us discuss the second case $\left(1-\delta\right)\rho n\le w_1,w_2\le\rho n$. Consider the distribution $\mathfrak{D}_2$ described as follows:
\begin{enumerate}
\item Sample $\cu:=\Pi_{i}^{\ell}\cu_i$ uniformly at random from $\cd_{\ell}(w,\mathbb{F}_q^{\eta\ell})$.
\item For each $i\in\left[1:\ell\right]$ sample $m$ vectors uniformly and independently at random from $\cu_i$ and set them as the rows of a matrix $X_i$.
\item Put $X_1,X_2,\ldots,X_\ell$ together to form an ordered $\ell$-tuple $\bm{X}:=\left(X_1,X_2,\ldots,X_\ell\right)$ of $\left(m\times\eta\right)$-matrices.
\end{enumerate}
Note that matrices $X_1,X_2,\ldots,X_\ell$ in the second step are constructed independently, since $\cu$ and hence $\cu_i$ for each $i\in\left[1:\ell\right]$ is determined after the first step. Thus, under the distribution $\mathfrak{D}_2$ we obtain an ordered $\ell$-tuple of $\left(m\times\eta\right)$-matrices of matrix sum-rank weight at most of $w$ with probability at least:
\begin{equation*}
\prod_{i=1}^{\ell}\left(\prod_{k=0}^{w_{i}-1}\frac{q^{w_{i}}-q^{k}}{q^{w_{i}^2}}\right)\ge\prod_{i=1}^{\ell}\left(\prod_{k=1}^{\infty}\left(1-q^{-k}\right)\right)=K_q^{\ell}.
\end{equation*}
Let $\bm{X}_1$ and $\bm{X}_2$ be independently sampled according to $\mathfrak{D}_2$. Then $\bff{Pr}_{\bm{X}_1,\bm{X}_2\sim\mathfrak{D}_2}\left[\ce\left(w_1,w_2\right)\right]\ge K_q^{2\ell}$, and hence:
\begin{equation}\label{eq:18}
\pr_{\bm{X}_1,\bm{X}_2\sim\mathfrak{D}_2}\left[\bm{X}_1+\bm{X}_2\in\cb_{\srk,\ell}\left(\bm{Y},\rho n\right)\vert\ce\left(w_1,w_2\right)\right]\le K_q^{-2\ell}\pr_{\bm{X}_1,\bm{X}_2\sim\mathfrak{D}_2}\left[\bm{X}_1+\bm{X}_2\in\cb_{\srk,\ell}\left(\bm{Y},\rho n\right)\right].
\end{equation}
Note that conditioned on the event $\ce\left(w_1,w_2\right)$, the distribution $\mathfrak{D}_1$ and $\mathfrak{D}_2$ are the same, that is, $\mathfrak{D}_1\vert\ce\left(w_1,w_2\right)$ and $\mathfrak{D}_2\vert\ce\left(w_1,w_2\right)$ are both uniform distributions over the pairs $\left(\bm{X}_1,\bm{X}_2\right)$ with $\wt_{\srk,\ell}\left(\bm{X}_j\right)=w_j$ for each $j\in\left[1:2\right]$. Then by \cref{eq:15} we have:
\begin{equation}\label{eq:19}
\begin{array}{ll}
\displaystyle{p} & \displaystyle{\le n^2\pr_{\bm{X}_1,\bm{X}_2\sim\mathfrak{D}_1}\left[\bm{X}_1+\bm{X}_2\in\cb_{\srk,\ell}\left(\bm{Y},\rho n\right)\vert \ce\left(w_1,w_2\right)\right]} \vspace{1ex}\\
\displaystyle{} & \displaystyle{=n^2\pr_{\bm{X}_1,\bm{X}_2\sim\mathfrak{D}_2}\left[\bm{X}_1+\bm{X}_2\in\cb_{\srk,\ell}\left(\bm{Y},\rho n\right)\vert \ce\left(w_1,w_2\right)\right].} 
\end{array}
\end{equation}
By combining \cref{eq:18} and \cref{eq:19} we obtain:
\begin{equation*}
p\le K_q^{-2\ell}n^2\pr_{\bm{X}_1,\bm{X}_2\sim\mathfrak{D}_2}\left[\bm{X}_1+\bm{X}_2\in\cb_{\srk,\ell}\left(\bm{Y},\rho n\right)\right].
\end{equation*} 
Thus, it is sufficient to show that:
\begin{equation*}
\pr_{\bm{X}_1,\bm{X}_2\sim\mathfrak{D}_2}\left[\bm{X}_1+\bm{X}_2\in\cb_{\srk,\ell}\left(\bm{Y},\rho n\right)\right]\le q^{-\Omega\left(mn\right)}.
\end{equation*}
Since it will finally lead to:
\begin{equation*}
p = \pr_{\bm{X}_1,\bm{X}_2\sim\mathfrak{D}_1}\left[\bm{X}_1+\bm{X}_2\in\cb_{\srk,\ell}\left(\bm{Y},\rho n\right)\right] \le K_q^{-2\ell}n^2 q^{-\Omega\left(mn\right)}=\left(1/q\right)^{\Omega\left(mn\right)-2\log_q\left(n\right)+2\ell\log_q(K_q)}=q^{-\Omega\left(mn\right)}.
\end{equation*}
Let $\cw_1$ and $\cw_2$ be $\ell$-decomposable subspaces of dimension $w_1$ and $w_2$ sampled independently according to the distribution $\mathfrak{D}_2$. Without loss of generality assume that $w_1\ge w_2$. Let $\epsilon\in\left(0,1\right)$. Setting $\alpha=\rho:=1-\epsilon$ in the dimension lemma we obtain the probability that
\begin{equation*}
\dim\left(\cw_1+\cw_2\right)\le w_1+\left(1-\alpha\right)w_2\le\left(1-\epsilon\right)n+\epsilon\left(1-\epsilon\right)n=(1-\epsilon^2)n
\end{equation*}
is at most $q^{-\Omega\left(\eta n\right)}=q^{-\Omega\left(mn\right)}$ for $m=\Theta\left(\eta\right)$. Now, we define the event $\cf$ to be $\dim\left(\cw_1+\cw_2\right)>(1-\epsilon^2)n$. Then $\bff{Pr}\left[\neg\cf\right]=q^{-\Omega\left(mn\right)}$. Using the law of total probability yields:
\begin{equation*}
\pr_{\bm{X}_1,\bm{X}_2\sim\mathfrak{D}_2}\left[\bm{X}_1+\bm{X}_2\in\cb_{\srk,\ell}\left(\bm{Y},\rho n\right)\right]\le q^{-\Omega\left(mn\right)}+\pr_{\bm{X}_1,\bm{X}_2\sim\mathfrak{D}_2}\left[\bm{X}_1+\bm{X}_2\in\cb_{\srk,\ell}\left(\bm{Y},\rho n\right)\vert\cf\right].
\end{equation*}
Now it is sufficient to prove:
\begin{equation}\label{eq:20}
\pr_{\bm{X}_1,\bm{X}_2\sim\mathfrak{D}_2}\left[\bm{X}_1+\bm{X}_2\in\cb_{\srk,\ell}\left(\bm{Y},\rho n\right)\vert\cf\right]=q^{-\Omega\left(mn\right)}.
\end{equation}
Recall that Lemma \ref{lem:3}.2 ensures that $\cw:=\cw_1+\cw_2$ is also an $\ell$-decomposable subspace, that is, if $\cw_j:=\Pi_{i}^{\ell}\cw_{j;i}$ for each $j\in\left[1:2\right]$, we have $\cw=\Pi_{i}^{\ell}\left(\cw_{1;i}+\cw_{2;i}\right)$. Thus, sampling $\bm{v}\in\cw$ uniformly at random is equivalent to sampling $v_i\in\cw_{1;i}+\cw_{2;i}$ independently and uniformly at random for all $i\in\left[1:\ell\right]$. And sampling $m$ vectors from $\cw$ independently and uniformly is equivalent to independently and uniformly sampling them from $\cw_{1;1}+\cw_{2;1},\cw_{1;2}+\cw_{2;2},\ldots,\cw_{1;\ell}+\cw_{2;\ell}$. Thus, conditioned on the event $\mathcal{F}$, which restricts the sample space, the probability of sampling $\bm{Z}\in\mathcal{M}^\ell$, where $\bm{Z}$ is the sum of two matrix tuples sampled according to $\mathfrak{D}_2$, is at most:
\begin{equation*}
\left(q^{-\dim\left(\cw\right)}\right)^m<\left(q^{-(1-\epsilon^2)n}\right)^m=q^{-(1-\epsilon^2)mn}.
\end{equation*}

Which gives us:
\begin{align*}
    \pr_{\bm{X}_1,\bm{X}_2\sim\mathfrak{D}_2}\left[\bm{X}_1+\bm{X}_2\in\cb_{\srk,\ell}\left(\bm{Y},\rho n\right)\vert\cf\right]  &\leq 
    \frac{\left|\cb_{\srk,\ell}\left(\bm{0},\rho n\right)\right|}{q^{\dim\left(\cw\right)}}\\[1.0ex]
    &< \frac{\left|\cb_{\srk,\ell}\left(\bm{0},\rho n\right)\right|}{q^{(1-\epsilon^2)mn}}\\[1.0ex]
    &\leq q^{-(1-\epsilon^2)mn}K_q^{-\ell}\binom{\ell+\left(1-\epsilon\right)n}{\ell}q^{\left(m+\epsilon \eta\right)\left(1-\epsilon\right)n}\\[1.0ex]
    &= q^{-\left(1-\epsilon\right)\epsilon\left(1-b\right)mn-\ell\log_q(K_q)+\log_{q} \binom{\ell+\left(1-\epsilon\right)n}{\ell}}.
\end{align*}

Define $T_2$ to be the negative exponent of $q$ in the equation above. Note that $\left(1-\epsilon\right)\epsilon\left(1-b\right)\ge 0$ for all $\epsilon\in\left(0,1\right)$ and $b\in\left(0,1\right]$. Thus, for sufficiently large $m$, $\eta$, and $\ell$ the cubic order term $\left(1-\epsilon\right)\epsilon\left(1-b\right)mn$ in $T_2$ dominates. Thus, one can find a positive constant $C'$ such that $T_2\ge C'mn$, which implies that $T_2=\Omega\left(mn\right)$ and concludes our proof.
\hfill$\square$
\end{pro}

By using the so-called $q$-\textit{ary} $c$-\textit{increasing chain lemma} \cite{Gur2010} one can generalize Theorem \ref{th:4} and obtain Lemma \ref{lem:5}.
\begin{de}[$q$-ary $c$-Increasing Chain \cite{Gur2010}]
\upshape
Let $c$ be an integer. A sequence of vectors $v_1,v_2,\ldots,v_d\in\mathbb{F}_q^{\kappa}$ is a $q$-\textit{ary} $c$-\textit{increasing chain of length} $d$, or simply $c$-\textit{increasing chain}, if for all $j\in\left[1:d\right]$:
\begin{equation*}
\left|\supp\left(v_j\right)\setminus\bigcup_{i=1}^{j-1}\supp\left(v_i\right)\right|\ge c,
\end{equation*}
where the \textit{support} of a vector $v\in\mathbb{F}_q^{\gamma}$ is defined to be $\supp\left(v\right):=\{k\in\left[1:\gamma\right]:v_k\ne 0_{\mathbb{F}_q}\}$.
\end{de}
\begin{lemma}[$q$-ary $c$-Increasing Chain Lemma \cite{Gur2010}]\label{lem:4}
For every prime power $q$, and all positive integers $c$, $\gamma$, and $A\le q^{\gamma}$, the
following holds: For every $\ca\subseteq\mathbb{F}_q^{\gamma}$ with $\left|\ca\right|=A$, there is a $w\in\mathbb{F}_q^{\gamma}$ such that $\ca+w$ has a $c$-increasing chain of length at least:
\begin{equation*}
\frac{1}{c}\log_q\left(\frac{A}{2}\right)-\left(1-\frac{1}{c}\right)\log_q\left(\left(q-1\right)\gamma\right).
\end{equation*}
\end{lemma}

From this point onward, the proof framework for all subsequent results follows the framework outlined in \cite{Gur2010,Din2014,Gur2017}. Lemma \ref{lem:5} generalizes the corresponding results of Hamming metric codes \cite[Claim 8]{Gur2010} and rank metric codes \cite[Claim 2]{Gur2017}. 
\begin{lemma}\label{lem:5}
Let $m$, $\eta$, and $\ell$ be positive integers. Let $\rho\in\left(0,1\right)$. Let $\bm{X}_1,\bm{X}_2,\ldots,\bm{X}_{\gamma}$ be sampled independently and uniformly at random from $\cb_{\srk,\ell}\left(\bm{0},\rho n\right)$. Let $A:=L+1:=K\gamma+1$ for some $K>1$ and $c=2$. Then for any $\ca\subseteq\mathbb{F}_q^{\gamma}$ of size $\left|\ca\right|=A$:
\begin{equation*}
\pr\left[\bigcap_{a\in\ca}\left(\sum_{i=1}^{\gamma}a_i\bm{X}_i\in\cb_{\srk,\ell}\left(\bm{0},\rho n\right)\right)\right]\le q^{-\left(dC-1\right)mn},
\end{equation*}
where $d$ satisfies:
\begin{equation*}
d>\frac{1}{2}\log_q\left(\frac{K}{2\left(q-1\right)q^2}\right)
\end{equation*}
is an integer, and $C$ is the constant in $\Omega\left(mn\right)$ determined in Theorem \ref{th:4}.
\end{lemma}
\begin{pro}
Let $\ca\subseteq\mathbb{F}_q^{\gamma}$ of size $\left|\ca\right|=A$. Note that $A\le q^{\gamma}$ for sufficiently large $\gamma$. Then by Lemma \ref{lem:4} we know that there is a $w\in\mathbb{F}_{q}^{\gamma}$ such that $\ca+w$ has a $c$-increasing chain of length $d$ with $d$ satisfying:
\begin{equation*}
\begin{array}{ll}
\displaystyle{d} & \displaystyle{=\left\lceil\frac{1}{c}\log_q\left(\frac{A}{2}\right)-\left(1-\frac{1}{c}\right)\log_q\left(\left(q-1\right)\gamma\right)\right\rceil\ge\frac{1}{2}\log_q\left(\frac{A}{2}\right)-\left(1-\frac{1}{2}\right)\log_q\left(\left(q-1\right)\gamma\right)} \vspace{1ex}\\
\displaystyle{} & \displaystyle{>\frac{1}{2}\left(\log_q\left(\frac{A-1}{2}\right)-\log_q\left(\left(q-1\right)\gamma\right)\right)-1=\frac{1}{2}\log_q\left(\frac{K}{2\left(q-1\right)q^2}\right).}
\end{array}
\end{equation*}
Let $v_1,v_2,\ldots,v_d\in\ca$ such that $v_1+w,v_2+w,\ldots,v_d+w$ is a $2$-increasing chain of length $d$. Then the monotonicity of probablity measure gives:
\begin{equation*}
\begin{array}{ll}
\displaystyle{} & \displaystyle{\pr\left[\bigcap_{a\in\ca}\left(\sum_{i=1}^{\gamma}a_i\bm{X}_i\in\cb_{\srk,\ell}\left(\bm{0},\rho n\right)\right)\right]{\le}\pr\left[\bigcap_{j=1}^{d}\left(\sum_{i=1}^{\gamma}v_{j;i}\bm{X}_i\in\cb_{\srk,\ell}\left(\bm{0},\rho n\right)\right)\right]} \vspace{1ex}\\
\displaystyle{\overset{\left(1\right)}{=}} & \displaystyle{\pr\left[\bigcap_{j=1}^{d}\left(\sum_{i=1}^{\gamma}\left(v_{j;i}+w_i\right)\bm{X}_i\in\cb_{\srk,\ell}\left(\sum_{i=1}^{\gamma}w_i\bm{X}_i,\rho n\right)\right)\right],}
\end{array}
\end{equation*}
where in equality $\left(1\right)$ we use Corollary \ref{cor:1}. Fix $\bm{Y}\in\cm^{\ell}$. Then:
\begin{equation}\label{eq:22}
\begin{array}{ll}
\displaystyle{} & \displaystyle{\pr\left[\bigcap_{j=1}^{d}\left(\sum_{i=1}^{\gamma}\left(v_{j;i}+w_i\right)\bm{X}_i\in\cb_{\srk,\ell}\left(\bm{Y},\rho n\right)\right)\right]} \vspace{1ex}\\
\displaystyle{{=}} & \displaystyle{\prod_{j=1}^{d}\pr\left[\left.\sum_{i=1}^{\gamma}\left(v_{j;i}+w_i\right)\bm{X}_i\in\cb_{\srk,\ell}\left(\bm{Y},\rho n\right)\right\vert\bigcap_{k=1}^{j-1}\left(\sum_{i=1}^{\gamma}\left(v_{k;i}+w_i\right)\bm{X}_i\in\cb_{\srk,\ell}\left(\bm{Y},\rho n\right)\right)\right],}
\end{array}
\end{equation}
Observe that for all $j\in\left[1:d\right]$ we have:
\begin{equation*}
\sum_{i=1}^{\gamma}\left(v_{j;i}+w_i\right)\bm{X}_i=\sum_{i=1}^{\gamma}\left(v_j+w\right)_i\bm{X}_i=\sum_{i\in\supp\left(v_j+w\right)}\left(v_j+w\right)_i\bm{X}_i=\sum_{i\in\supp\left(v_j+w\right)}\left(v_{j;i}+w_i\right)\bm{X}_i
\end{equation*}
Thus, the probability in \cref{eq:22} can be equivalently written as:
\begin{equation}\label{eq:23}
\prod_{j=1}^{d}\pr\left[\left.\sum_{i\in\supp\left(v_j+w\right)}\left(v_{j;i}+w_i\right)\bm{X}_i\in\cb_{\srk,\ell}\left(\bm{Y},\rho n\right)\right\vert\bigcap_{k=1}^{j-1}\left(\sum_{i\in\supp\left(v_k+w\right)}\left(v_{k;i}+w_i\right)\bm{X}_i\in\cb_{\srk,\ell}\left(\bm{Y},\rho n\right)\right)\right].
\end{equation}
Now for all $k\in\left[1:j-1\right]$ and for all $i\in\supp\left(v_k+w\right)$, that is, for all $t\in\bigcup_{k=1}^{j-1}\supp\left(v_k+w\right)$ we fix $\bm{X}_t$ such that the probability that:
\begin{equation}\label{eq:24}
\sum_{i\in\supp\left(v_j+w\right)}\left(v_{j;i}+w_i\right)\bm{X}_i\in\cb_{\srk,\ell}\left(\bm{Y},\rho n\right)
\end{equation}
as large as possible. Thus, the conditional probability in \cref{eq:23} is at most:
\begin{equation}\label{eq:25}
\prod_{j=1}^{d}\max_{
\begin{smallmatrix}
\bm{Z}_t\in\cb_{\srk,\ell}\left(\bm{0},\rho n\right)\\
t\in\bigcup_{k=1}^{j-1}\supp\left(v_k+w\right)\end{smallmatrix}}
\pr\left[\left.\sum_{i\in\supp\left(v_j+w\right)}\left(v_{j;i}+w_i\right)\bm{X}_i\in\cb_{\srk,\ell}\left(\bm{Y},\rho n\right)\right\vert\bigcap_{t\in\bigcup_{k=1}^{j-1}\supp\left(v_k+w\right)}\left(\bm{X}_t=\bm{Z}_t\right)\right],
\end{equation}
where $\bm{Z}_t$ are fixed matrix tuples in $\cb_{\srk,\ell}\left(\bm{0},\rho n\right)$ for all $t\in\bigcup_{k=1}^{j-1}\supp\left(v_k+w\right)$. Since $v_1+w,v_2+w,\ldots,v_d+w$ is a $2$-increasing chain, we can choose distinct elements $i_1,i_2$ in $\supp\left(v_j+w\right)\setminus\bigcup_{k=1}^{j-1}\supp\left(v_k+w\right)$ for all $j\in\left[1:d\right]$\footnote{Here without loss of generality we can assume that $|\supp\left(v_j+w\right)\setminus\bigcup_{k=1}^{j-1}\supp\left(v_k+w\right)|$ is exactly equal to $2$ for all $j\in\left[1:d\right]$.}. Then we have:
\begin{equation}\label{eq:26}
\sum_{i\in\supp\left(v_j+w\right)}\left(v_{j;i}+w_i\right)\bm{X}_i=\left(v_{j;i_1}+w_{i_1}\right)\bm{X}_{i_1}+\left(v_{j;i_2}+w_{i_2}\right)\bm{X}_{i_2}+\sum_{i\in\supp\left(v_j+w\right)\setminus\left\{i_1,i_2\right\}}\left(v_{j;i}+w_i\right)\bm{Z}_i.
\end{equation}
By combining Corollary \ref{cor:1} and \cref{eq:26} the event in \cref{eq:24} can be equivalently written as:
\begin{equation*}
\left(v_{j;i_1}+w_{i_1}\right)\bm{X}_{i_1}+\left(v_{j;i_2}+w_{i_2}\right)\bm{X}_{i_2}\in\cb_{\srk,\ell}\left(\bm{Y}-\sum_{i\in\supp\left(v_j+w\right)\setminus\left\{i_1,i_2\right\}}\left(v_{j;i}+w_i\right)\bm{Z}_i,\rho n\right)
\end{equation*}
for all $j\in\left[1:d\right]$. For each $j\in\left[1:d\right]$ define:
\begin{equation*}
\begin{array}{l}
\displaystyle{\hat{\bm{X}}_{j;1}:=\left(v_{j;i_1}+w_{i_1}\right)\bm{X}_{i_1},} \vspace{1ex}\\
\displaystyle{\hat{\bm{X}}_{j;2}:=\left(v_{j;i_2}+w_{i_2}\right)\bm{X}_{i_2},} \vspace{1ex}\\
\displaystyle{\hat{\bm{Y}}_j:=\bm{Y}-\sum_{i\in\supp\left(v_j+w\right)\setminus\left\{i_1,i_2\right\}}\left(v_{j;i}+w_i\right)\bm{Z}_i,} 
\end{array}
\end{equation*}
where $\hat{\bm{Y}}_j$ is fixed for each $j\in\left[1:d\right]$. Note also that by Corollary \ref{cor:1} we have:
\begin{equation*}
(\hat{\bm{X}}_{j;k}\in\cb_{\srk,\ell}\left(\bm{0},\rho n\right))\Leftrightarrow\left(\bm{X}_{i_k}\in\cb_{\srk,\ell}\left(\bm{0},\rho n\right)\right)
\end{equation*}
for all $k\in\left[1:2\right]$. Thus, since $\bm{X}_{i_1}$ and $\bm{X}_{i_2}$ are independent and uniform samples over $\cb_{\srk,\ell}\left(\bm{0},\rho n\right)$, so are $\hat{\bm{X}}_{j;1}$ and $\hat{\bm{X}}_{j;2}$. Finally, Theorem \ref{th:4} can be applied to $\hat{\bm{X}}_{j;1}$, $\hat{\bm{X}}_{j;2}$, and $\hat{\bm{Y}}_j$ for all $j\in\left[1:d\right]$, which gives an upper bound of the probability in \cref{eq:25}:
\begin{equation*}
\begin{array}{ll}
\displaystyle{} & \displaystyle{\prod_{j=1}^{d}\max_{
\begin{smallmatrix}
\bm{Z}_t\in\cb_{\srk,\ell}\left(\bm{0},\rho n\right)\\
t\in\bigcup_{k=1}^{j-1}\supp\left(v_k+w\right)\end{smallmatrix}}
\pr\left[\left.\sum_{i\in\supp\left(v_j+w\right)}\left(v_{j;i}+w_i\right)\bm{X}_i\in\cb_{\srk,\ell}\left(\bm{Y},\rho n\right)\right\vert\bigcap_{t\in\bigcup_{k=1}^{j-1}\supp\left(v_k+w\right)}\left(\bm{X}_t=\bm{Z}_t\right)\right]} \vspace{1ex}\\
\displaystyle{\le} & \displaystyle{q^{-dCmn},} 
\end{array}
\end{equation*}
where $C$ is the constant in $\Omega\left(mn\right)$ determined in Theorem \ref{th:4}. Finally, taking union bound on all possible $\left(q^{m\eta}\right)^{\ell}=q^{mn}$ choices of $\bm{Y}\in\cm^{\ell}$ gives the desired result.
\hfill$\square$
\end{pro}

Now we will give the formal statement of the limited correlation property between sum-rank metric balls and subspaces and prove it. This result generalizes the limited correlation properties between Hamming metric balls and subspaces \cite[Theorem 3]{Gur2010} and between rank metric balls and subspaces \cite[Lemma 4.2]{Gur2017}.

\begin{lemma}[Limited Correlation Property]\label{lem:6}
For every $\rho\in\left(0,1\right)$ there is a constant $K>1$ such that for all integers $m$, $\eta$, and $\ell$, and $\gamma=o(\sqrt{mn})$, if $\bm{X}_1,\bm{X}_2,\ldots,\bm{X}_{\gamma}$ be sampled uniformly and independently at random from $\cb_{\srk,\ell}\left(\bm{0},\rho n\right)$, then:
\begin{equation*}
\pr[|\spann\left(\left\{\bm{X}_1,\bm{X}_2,\ldots,\bm{X}_{\gamma}\right\}\right)\cap\cb_{\srk,\ell}\left(\bm{0},\rho n\right)|\ge K\gamma]\le q^{-\left(3-o\left(1\right)\right)mn}.
\end{equation*}
\end{lemma}
\begin{pro}
To provide an upper bound of the probability of interest, we would like to find an upper bound for the probability that more than $L:=K\gamma$ elements in $\spann\left(\left\{\bm{X}_1,\bm{X}_2,\ldots,\bm{X}_{\gamma}\right\}\right)$ lie in $\cb_{\srk,\ell}\left(\bm{0},\rho n\right)$. Note that this event occurs if and only if there exists a subset $\ca\subseteq\mathbb{F}_q^{\gamma}$ of size $\left|\ca\right|=L+1$ such that:
\begin{equation*}
\sum_{i=1}^{\gamma} a_i\bm{X}_i\in\cb_{\srk,\ell}\left(\bm{0},\rho n\right),\quad \forall a\in\ca.
\end{equation*}
By Lemma \ref{lem:5} we know that the probability of this event is at most $q^{-\left(dC-1\right)mn}$, where $d$ and $C$ are determined in Lemma \ref{lem:5}. Taking the union bound on all possible choices of $\ca$ yields:
\begin{equation}\label{eq:27}
\sum_{\substack{\ca\subseteq\mathbb{F}_q^{\gamma}\\\left|\ca\right|=L+1}}q^{-\left(dC-1\right)mn}=\binom{q^{\gamma}}{L+1}q^{-\left(dC-1\right)mn}\le q^{\gamma\left(K\gamma+1\right)}q^{-\left(dC-1\right)mn}=q^{o\left(mn\right)-\left(dC-1\right)mn}.
\end{equation}
Assuming $K$ sufficiently large such that $d\ge 4/C$ we find that the probability in \cref{eq:27} is at most $q^{-\left(3-o\left(1\right)\right)mn}$.
\hfill$\square$
\end{pro}

Finally, Theorem \ref{th:5} states that any random $\mathbb{F}_q$-linear sum-rank metric code of rate not exceeding the list-decoding capacity is $\left(\rho,O\left(1/\epsilon\right)\right)$-list-decodable with high probability.
\begin{theorem}[List-Decodability for Random $\mathbb{F}_q$-Linear Sum-Rank Metric Codes]\label{th:5}
Let $m$, $\eta$, and $\ell$ be positive integers with $\eta\le m$. Define $b:=\eta/m$. Then there exists a constant $C>0$ such that a random $\mathbb{F}_q$-linear sum-rank metric code $\cc\le\cm^{\ell}$ of rate $R=1-\kappa_b\left(\rho\right)-\epsilon$ is $\left(\rho,C/\epsilon\right)$-list-decodable with high probability for $\rho\in\left(0,1\right)$, that is:
\begin{equation*}
\pr\left[\cc\,\,\text{is}\,\,\left(\rho,O\left(1/\epsilon\right)\right)\text{-}\text{list\text{-}decodable}\right]\ge 1-q^{-\Omega\left(mn\right)}.
\end{equation*}
\end{theorem}
\begin{pro}
Let $L=\left\lceil 2K/\epsilon\right\rceil$, where $K$ is the constant from Lemma \ref{lem:6}, and let $m$, $\eta$ and $\ell$ be sufficiently large compared to $1/\epsilon$ such that the $o\left(1\right)$ term in Lemma \ref{lem:6} is at most $1$. Let $\bm{X}\in\cm^{\ell}$ be sampled uniformly at random. We will study the following probability:
\begin{equation*}
p:=\pr_{\bm{X},\cc}\left[\left|\cb_{\srk,\ell}\left(\bm{X},\rho n\right)\cap \cc\right|\ge L\right].
\end{equation*}
Note that:
\begin{align}\label{eq:28}
    \begin{split}
    p = \pr_{\bm{X},\cc}\left[\left|\cb_{\srk,\ell}\left(\bm{X},\rho n\right)\cap \cc\right|\ge L\right]
    & \overset{\left(1\right)}{=}\pr_{\bm{X},\cc}\left[\left|\cb_{\srk,\ell}\left(\bm{0},\rho n\right)\cap\left(\cc-\bm{X}\right)\right|\ge L\right]\\[1.0ex]
    & \overset{\left(2\right)}{\le}\pr_{\hat{\cc}}[|\cb_{\srk,\ell}\left(\bm{0},\rho n\right)\cap\hat{\cc}|\ge L],
    \end{split}
\end{align}

\noindent where in equality $\left(1\right)$ we use the fact that $\left|\cb_{\srk,\ell}\left(\bm{X},\rho n\right)\cap \cc\right|=\left|\cb_{\srk,\ell}\left(\bm{0},\rho n\right)\cap\left(\cc-\bm{X}\right)\right|$. To see this, let $\bm{Y}\in\cb_{\srk,\ell}\left(\bm{X},\rho n\right)\cap \cc$ and hence $\bm{Y}\in\cc$ and $\bm{Y}\in\cb_{\srk,\ell}\left(\bm{X},\rho n\right)$. By Corollary \ref{cor:1} we know that $\bm{Y}$ can be written as $\bm{X}+\bm{Z}$ for some $\bm{Z}\in\cb_{\srk,\ell}\left(\bm{0},\rho n\right)$. Moreover, we note that $\bm{Z}=\bm{Y}-\bm{X}\in\left(\cc-\bm{X}\right)$, and hence $\bm{Z}\in\cb_{\srk,\ell}\left(\bm{0},\rho n\right)\cap\left(\cc-\bm{X}\right)$. Now let us define a map $\tau$ from $\cc$ to the affine space $\cc-\bm{X}$ by:
\begin{equation*}
\tau:\cc\to\cc-\bm{X},\quad \bm{Y}\mapsto\tau\left(\bm{Y}\right)=\bm{Y}-\bm{X}=:\bm{Z}.
\end{equation*}
$\tau$ is a translation and thus bijective. Thus, we proved that $\left|\cb_{\srk,\ell}\left(\bm{X},\rho n\right)\cap \cc\right|=\left|\cb_{\srk,\ell}\left(\bm{0},\rho n\right)\cap\left(\cc-\bm{X}\right)\right|$. In inequality $\left(2\right)$ of \cref{eq:28} we define
\begin{equation*}
\hat{\cc}:=\left\{
\begin{array}{ll}
\displaystyle{\cc+\spann\left(\left\{\bm{X}\right\}\right),} & \displaystyle{\bm{X}\notin\cc,} \vspace{1ex}\\
\displaystyle{\cc+\spann\left(\left\{\bm{X}'\right\}\right),} & \displaystyle{\bm{X}\in\cc,\bm{X}'\in\cm^{\ell}\setminus\cc}  
\end{array}
\right.
\end{equation*}
to be a random $(Rmn+1)$-dimensional subspace. If $\bm{X}\notin\cc$, let $\bm{Y}-\bm{X}\in\cc-\bm{X}$. Then we have $\bm{Y}-\bm{X}\in\cc+(-1_{\mathbb{F}_q})\bm{X}\subseteq\cc+\spann\left(\left\{\bm{X}\right\}\right)$. Thus, we have $\cc-\bm{X}\subseteq\hat{\cc}$ for $\bm{X}\notin\cc$\footnote{Note that here we actually discuss the set inclusion, since the affine space $\cc-\bm{X}$ is \textit{not} a subspace in general.}. Similarly, if $\bm{X}\in\cc$, let $\bm{Y}-\bm{X}\in\cc-\bm{X}=\cc$. Then $\bm{Y}-\bm{X}\in\cc=\cc+\left\{\bm{0}\right\}\subseteq\cc+\spann\left(\left\{\bm{X}'\right\}\right)$ for some $\bm{X}'\in\cm^{\ell}\setminus\cc$. Thus, we have $\cc-\bm{X}\subseteq\hat{\cc}$ for $\bm{X}\in\cc$ as well. For each $\gamma\in[\lceil\log_q\left(L\right)\rceil:L]$ define the set $\ct_{\gamma}\cb$ to be the set of all $\gamma$-tuples $\left(\bm{M}_1,\bm{M}_2,\ldots,\bm{M}_{\gamma}\right)\in\cb^{\gamma}_{\srk,\ell}\left(\bm{0},\rho n\right)$ such that $\bm{M}_1,\bm{M}_2,\ldots,\bm{M}_{\gamma}$ are linearly independent and $|\spann\left(\left\{\bm{M}_1,\bm{M}_2,\ldots,\bm{M}_{\gamma}\right\}\right)\cap\cb_{\srk,\ell}\left(\bm{0},\rho n\right)|\ge L$. Define further:
\begin{equation*}
\ct\cb:=\bigcup_{\gamma\in[\lceil\log_q\left(L\right)\rceil:L]}\ct_{\gamma}\cb.
\end{equation*}
Note that if the event $\{|\cb_{\srk,\ell}\left(\bm{0},\rho n\right)\cap\hat{\cc}|\ge L\}$ occurs, then there must exist some $\left(\bm{M}_1,\bm{M}_{2},\ldots,\bm{M}_{\gamma}\right)\in\ct\cb$ for some $\gamma$ for which $\left\{\bm{M}_1,\bm{M}_2,\ldots,\bm{M}_{\gamma}\right\}\subseteq\hat{\cc}$. Thus, we have:
\begin{equation}\label{eq:29}
\begin{array}{ll}
\displaystyle{p} & \displaystyle{\le\pr_{\hat{\cc}}[|\cb_{\srk,\ell}\left(\bm{0},\rho n\right)\cap\hat{\cc}|\ge L]} \vspace{1ex}\\
\displaystyle{} & \displaystyle{{\le}\pr_{\hat{\cc}}[\exists\left(\bm{M}_1,\bm{M}_2,\ldots,\bm{M}_{\gamma}\right)\in\ct\cb,\gamma\in[\lceil\log_q\left(L\right)\rceil:L]:\left\{\bm{M}_1,\bm{M}_2,\ldots,\bm{M}_{\gamma}\right\}\subseteq\hat{\cc}]} \vspace{1ex}\\
\displaystyle{} & \displaystyle{{\le}\sum_{\gamma=\lceil\log_q\left(L\right)\rceil}^{L}\sum_{\left(\bm{M}_1,\bm{M}_2,\ldots,\bm{M}_{\gamma}\right)\in\ct_{\gamma}\cb}\pr_{\hat{\cc}}[\left\{\bm{M}_1,\bm{M}_2,\ldots,\bm{M}_{\gamma}\right\}\subseteq\hat{\cc}],} 
\end{array}
\end{equation}
where:
\begin{equation}\label{eq:40}
\pr_{\hat{\cc}}[\left\{\bm{M}_1,\bm{M}_2,\ldots,\bm{M}_{\gamma}\right\}\subseteq\hat{\cc}]{=}\prod_{i=1}^{\gamma}\pr_{\hat{\cc}}\left[\bm{M}_i\in\hat{\cc}\left\vert\bigcap_{j=1}^{i-1}(\bm{M}_j\in\hat{\cc})\right.\right]=\prod_{i=1}^{\gamma}\frac{q^{Rmn+1}-q^{i-1}}{q^{mn}-q^{i-1}}\le q^{\left(R-1\right)\gamma mn+\gamma}
\end{equation}
by linear independence among $\bm{M}_1,\bm{M}_2,\ldots,\bm{M}_{\gamma}$. Therefore, by combining \cref{eq:29} and \cref{eq:40} we obtain:
\begin{equation*}
p\le\sum_{\gamma=\lceil\log_q\left(L\right)\rceil}^L\left|\ct_{\gamma}\cb\right|q^{\left(R-1\right)\gamma mn+\gamma}.
\end{equation*}
We now bound $\left|\ct_{\gamma}\cb\right|$ depending on the value of $\gamma$.
\begin{enumerate}
\item $\gamma\in[\lceil\log_q\left(L\right)\rceil:\left\lfloor 2/\epsilon\right\rfloor]$. In this case, note that $\left|\ct_{\gamma}\cb\right|/|\cb^{\gamma}_{\srk,\ell}\left(\bm{0},\rho n\right)|$ is a lower bound on the probability that $\bm{X}_1,\bm{X}_2,\ldots,\bm{X}_{\gamma}$ sampled independently and uniformly at random from $\cb_{\srk,\ell}\left(\bm{0},\rho n\right)$ are such that $|\spann\left(\left\{\bm{X}_1,\bm{X}_2,\ldots,\bm{X}_{\gamma}\right\}\right)\cap\cb_{\srk,\ell}\left(\bm{0},\rho n\right)|\ge L=\left\lceil 2K/\epsilon\right\rceil$. Since $\gamma\le\left\lfloor 2/\epsilon\right\rfloor\le 2/\epsilon+1\le 2/\epsilon^{3/2}+1\le 2\sqrt{m\eta\ell}+1=2\sqrt{mn}+1$, then Lemma \ref{lem:6} can be applied and hence:
\begin{equation*}
\pr[|\spann\left(\left\{\bm{X}_1,\bm{X}_2,\ldots,\bm{X}_{\gamma}\right\}\right)\cap\cb_{\srk,\ell}\left(\bm{0},\rho n\right)|\ge K\gamma]\le q^{-2mn}.
\end{equation*}
Since $K\left\lfloor 2/\epsilon\right\rfloor\ge K\gamma$ for each $\gamma\in[\lceil\log_q\left(L\right)\rceil:\left\lceil 2/\epsilon\right\rceil]$, the monotonicity of probability measure yields:
\begin{equation*}
\pr[|\spann\left(\left\{\bm{X}_1,\bm{X}_2,\ldots,\bm{X}_{\gamma}\right\}\right)\cap\cb_{\srk,\ell}\left(\bm{0},\rho n\right)|\ge K\left\lceil 2/\epsilon\right\rceil]\le q^{-2mn}.
\end{equation*}
We also note that $L=\left\lceil 2K/\epsilon\right\rceil\ge K\left\lfloor 2/\epsilon\right\rfloor$. To see this, let $2/\epsilon:=a\in\mathbb{N}$ at first. In this case we have $L=Ka\ge Ka=K\left\lfloor 2/\epsilon\right\rfloor$. Then let $2/\epsilon:=a+\delta$, where $a\in\mathbb{N}$ and $\delta\in\left(0,1\right)$. Then we have $K\left\lfloor 2/\epsilon\right\rfloor=Ka$ and $\left\lceil 2K/\epsilon\right\rceil=\left\lceil Ka+K\delta\right\rceil\ge Ka+1$. Then in this case we also have $\left\lceil 2K/\epsilon\right\rceil\ge Ka+1>Ka=K\left\lfloor 2/\epsilon\right\rfloor$. By using the monotonicity of probability measure again it can be conclude that:
\begin{equation*}
\pr[|\spann\left(\left\{\bm{X}_1,\bm{X}_2,\ldots,\bm{X}_{\gamma}\right\}\right)\cap\cb_{\srk,\ell}\left(\bm{0},\rho n\right)|\ge \left\lceil 2K/\epsilon\right\rceil]\le q^{-2mn}.
\end{equation*}
Thus, we have:
\begin{equation*}
\left|\ct_{\gamma}\cb\right|\le|\cb^{\gamma}_{\srk,\ell}\left(\bm{0},\rho n\right)|q^{-2mn}<K_q^{-\ell\gamma}\binom{\ell+\rho n}{\ell}^{\gamma}q^{\left(m+\eta-\rho \eta\right)\rho\gamma n}q^{-2mn}.
\end{equation*}
\item $\gamma\in\left[\left\lfloor 2/\epsilon\right\rfloor+1:L\right]$. In this case, we note that $L=\left\lceil 2K/\epsilon\right\rceil\le K\gamma$ for all $\gamma\in\left[\left\lfloor 2/\epsilon\right\rfloor+1:L\right]$. To see this, let $2/\epsilon:=a\in\mathbb{N}$ at first. In this case we have $L=Ka<Ka+K=K\left(\left\lfloor 2/\epsilon\right\rfloor+1\right)\le K\gamma$. Then let $2/\epsilon:=a+\delta$, where $a\in\mathbb{N}$ and $\delta\in\left(0,1\right)$. Then we have $K\left(\left\lfloor 2/\epsilon\right\rfloor+1\right)=Ka+K$ and $\left\lceil 2K/\epsilon\right\rceil=\left\lceil Ka+K\delta\right\rceil\le Ka+K$. Then in this case we also have $L=\left\lceil 2K/\epsilon\right\rceil\le Ka+K=K\left(\left\lfloor 2/\epsilon\right\rfloor+1\right)\le K\gamma$. Thus, in this case we have:
\begin{equation*}
\left|\ct_{\gamma}\cb\right|\le|\cb^{\gamma}_{\srk,\ell}\left(\bm{0},\rho n\right)|<K_q^{-\ell\gamma}\binom{\ell+\rho n}{\ell}^{\gamma}q^{\left(m+\eta-\rho \eta\right)\rho\gamma n}.
\end{equation*}
\end{enumerate}
Combining these inequalities, we obtain the following upper bound of $p$:
\begin{equation}\label{eq:31}
p\le \sum_{\gamma=\lceil\log_q\left(L\right)\rceil}^{\lfloor 2/\epsilon\rfloor}\left|\ct_{\gamma}\cb\right|q^{\left(R-1\right)\gamma mn+\gamma}+\sum_{\gamma=\lfloor 2/\epsilon\rfloor+1}^{L}\left|\ct_{\gamma}\cb\right|q^{\left(R-1\right)\gamma mn+\gamma}=:p_{1}+p_2.
\end{equation}
We will bound $p_1$ and $p_2$ in \cref{eq:31} individually by using inequality scaling trick. For $p_1$ we have:
\begin{equation*}
\begin{array}{ll}
\displaystyle{p_1} & \displaystyle{=\sum_{\gamma=\lceil\log_q\left(L\right)\rceil}^{\lfloor 2/\epsilon\rfloor}\left|\ct_{\gamma}\cb\right|q^{\left(R-1\right)\gamma mn+\gamma}<\sum_{\gamma=\lceil\log_q\left(L\right)\rceil}^{\lfloor 2/\epsilon\rfloor}K_q^{-\ell\gamma}\binom{\ell+\rho n}{\ell}^{\gamma}q^{\left(m+\eta-\rho \eta\right)\rho\gamma n}q^{-2mn}q^{\left(R-1\right)\gamma mn+\gamma}} \vspace{1ex}\\
\displaystyle{} & \displaystyle{\le q^{-2mn}\sum_{\gamma=\lceil\log_q\left(L\right)\rceil}^{\lfloor 2/\epsilon\rfloor}\left(K_q^{-\ell}q\right)^{\gamma}\binom{\ell+\rho n}{\ell}^{\gamma}q^{(\rho+\rho b-\rho^2 b)\gamma mn}q^{(-\rho-\rho b+\rho^2b-\epsilon)\gamma mn}} \vspace{1ex}\\
\displaystyle{} & \displaystyle{=q^{-2mn}\sum_{\gamma=\lceil\log_q\left(L\right)\rceil}^{\lfloor 2/\epsilon\rfloor}\left(K_q^{-\ell}q\right)^{\gamma}\binom{\ell+\rho n}{\ell}^{\gamma}q^{-\epsilon\gamma mn}\le q^{-2mn}L\left(K_q^{-\ell}q\right)^L\binom{\ell+\rho n}{\ell}^{L}.} 
\end{array}
\end{equation*}
Next let us bound $p_2$ from above as follows:
\begin{equation*}
\begin{array}{ll}
\displaystyle{p_2} & \displaystyle{=\sum_{\gamma=\lfloor 2/\epsilon\rfloor+1}^{L}\left|\ct_{\gamma}\cb\right|q^{\left(R-1\right)\gamma mn+\gamma}<\sum_{\gamma=\lfloor 2/\epsilon\rfloor+1}^{L}K_q^{-\ell\gamma}\binom{\ell+\rho n}{\ell}^{\gamma}q^{\left(m+\eta-\rho\eta\right)\rho\gamma n}q^{\left(R-1\right)\gamma mn+\gamma}} \vspace{1ex}\\
\displaystyle{} & \displaystyle{\le\sum_{\gamma=\lfloor 2/\epsilon\rfloor+1}^{L}\left(K_q^{-\ell}q\right)^{\gamma}\binom{\ell+\rho n}{\ell}^{\gamma}q^{(\rho+\rho b-\rho^2 b)\gamma mn}q^{(-\rho-\rho b+\rho^2b-\epsilon)\gamma mn}} \vspace{1ex}\\
\displaystyle{} & \displaystyle{=\sum_{\gamma=\lfloor 2/\epsilon\rfloor+1}^{L}\left(K_q^{-\ell}q\right)^{\gamma}\binom{\ell+\rho n}{\ell}^{\gamma}q^{-\epsilon\gamma mn}\le L\left(K_q^{-\ell}q\right)^{L}\binom{\ell+\rho n}{\ell}^{L}q^{-2\epsilon mn/\epsilon}.} 
\end{array}
\end{equation*}
Thus:
\begin{equation*}
p=p_1+p_2\le 2L\left(K_q^{-\ell}q\right)^{L}\binom{\ell+\rho n}{\ell}^{L}q^{-2 mn}=q^{-2 mn+\log_q(2L)+L-\ell L\log_q(K_q)+L\log_q(C^{\ell+\rho n}_{\ell})}
\end{equation*}
Finally, by taking union bound on all possible choices of $\bm{X}\in\cm^{\ell}$, the probability that $\cc$ fails its $\left(\rho,L\right)$-list-decodability is at most:
\begin{equation*}
q^{mn}p\le q^{-mn+\log_q(2L)+L-\ell L\log_q(K_q)+L\log_q(C^{\ell+\rho n}_{\ell})}.
\end{equation*}
Thus, for sufficiently large $m$, $\eta$, and $\ell$ compared to $1/\epsilon$ the term $-mn$ in $q^{mn}p$ dominates. Finally, we have $q^{mn}p\le q^{-\Omega\left(mn\right)}$.
\hfill$\square$
\end{pro}

We observe that the list-decoding capacity is the key factor determining the list size. Therefore, we compare the list-decoding capacity of sum-rank metric codes with the list-decoding capacity of Hamming and rank metric codes. Based on our derivations, we first observe that the list-decoding capacity of sum-rank metric codes with block size $\left(m \times \eta\right)$ is the same as that of rank metric codes with code length $\eta$. This conclusion, although surprising, has its causal necessity, because when we review the proof of Theorem \ref{th:1}, we can see that the definition of random general sum-rank metric codes, the definition of list-decodability, and the upper bound of the volume of sum-rank metric balls lead to this inevitable result. On the other hand, when $m=\eta=1$, the sum-rank metric codes with $\ell$ blocks degenerate into Hamming metric codes with code length $\ell$. At this point, the list-decoding capacity of sum-rank metric codes is $1-\kappa_1\left(\rho\right)=1-\rho(2-\rho)$. We find that this value is very close to the list-decoding capacity of Hamming metric codes $1-H_q\left(\rho\right)$ when $q$ is small. However, as $q$ increases, $1-\kappa_1\left(\rho\right)$ will become smaller than $1-H_q\left(\rho\right)$, because as $q\to\infty$, the function $1-H_q\left(\rho\right)$ approaches $1-\rho$ \cite{Res2019}.

\section{Conclusions and Outlook}\label{sec:4}
In this paper, we investigate the list-decodability of random ($\mathbb{F}_q$-linear) sum-rank metric codes. Specifically, we begin by proving the list-decoding capacity theorem for sum-rank metric codes, which implies the list-decodability theorem for random general sum-rank metric codes. The result shows that any random general sum-rank metric code with a rate \textit{not} exceeding the list-decoding capacity $1-\kappa_b\left(\rho\right)$ is $\left(\rho,O\left(1/\epsilon\right)\right)$-list-decodable with high probability. 

Using the same proof techniques, we further establish a trivial version of list-decodability theorem for random $\mathbb{F}_q$-linear sum-rank metric codes. However, due to the high correlation among codewords in linear codes, the list size in this version is exponentially large. Therefore, we proceed to improve this exponentially large list size by leveraging the so-called limited correlation property between sum-rank metric balls and subspaces, which generalizes the corresponding results of Hamming and rank metrics. Finally, we complete the proof of the improved version of the list-decodability theorem for random $\mathbb{F}_q$-linear sum-rank metric codes. This result indicates that any random $\mathbb{F}_q$-linear sum-rank metric code with a rate \textit{not} exceeding the list-decoding capacity is $\left(\rho,O\left(1/\epsilon\right)\right)$-list-decodable with high probability. Ultimately, we achieve an exponential improvement on the list size. Our work then closes a gap in the analysis of list-decobability of random codes.

In the future, a potential open problem is the \textit{list-decodability of random} $\mathbb{F}_{q^m}$-\textit{linear vector sum}-\textit{rank metric codes}, i.e., uniform $\mathbb{F}_{q^m}$-subspaces of $\mathbb{F}_{q^m}^{\eta\ell}$. In fact, to the best of the author's knowledge, even though for random $\mathbb{F}_{q^m}$-linear rank metric codes, the same question, i.e., the list-decodability of random $\mathbb{F}_{q^m}$-linear rank metric codes, is still a research topic. The author's perspective in \cite{Res2019} suggests that for random $\mathbb{F}_{q^m}$-linear rank metric codes, their list sizes \textit{cannot} be polynomially large. Therefore, being able to prove or disprove whether random $\mathbb{F}_{q^m}$-linear vector sum-rank metric codes can have polynomially large list sizes would help us better understand their list-decodability.

\newpage
\appendix
\section{Appendix}\label{ap:1}

\begin{pro}[Proposition \ref{prop:2}]
Let $\bm{X},\bm{Y},\bm{Z}\in\cm_i^{\ell}$. Then:
\begin{equation*}
\begin{array}{ll}
\displaystyle{d_{\srk,\ell}\left(\bm{X}\pm \bm{Z},\bm{Y}\pm \bm{Z}\right)} & \displaystyle{=\wt_{\srk,\ell}\left(\left(\bm{X}\pm \bm{Z}\right)-\left(\bm{Y}\pm \bm{Z}\right)\right)=\wt_{\srk,\ell}\left(\bm{X}\pm \bm{Z}-\bm{Y}\mp \bm{Z}\right)} \vspace{1ex}\\
\displaystyle{} & \displaystyle{=\wt_{\srk,\ell}\left(\bm{X}-\bm{Y}\right)=d_{\srk,\ell}\left(\bm{X},\bm{Y}\right).}
\end{array}
\end{equation*}
Let $\alpha\in\mathbb{F}_q\setminus\{0_{\mathbb{F}_q}\}$. Then:
\begin{equation}\label{eq:39}
\begin{array}{ll}
\displaystyle{d_{\srk,\ell}\left(\alpha \bm{X},\alpha \bm{Y}\right)} & \displaystyle{=\sum_{i=1}^{\ell}\rk\left(\alpha X_i-\alpha Y_i\right)=\sum_{i=1}^{\ell}\rk\left(\alpha\left(X_i-Y_i\right)\right)} \vspace{1ex}\\
\displaystyle{} & \displaystyle{\overset{\left(1\right)}{=}\sum_{i=1}^{\ell}\rk\left(X_i-Y_i\right)=d_{\srk,\ell}\left(\bm{X},\bm{Y}\right),}
\end{array}
\end{equation}
where in equality $\left(1\right)$ in \cref{eq:39} we use the fact that multiplying a matrix by a non-zero scalar \textit{does not} change the rank of the matrix.
\hfill$\square$
\end{pro}

\begin{pro}[Corollary \ref{cor:1}]
Let $\bm{X},\bm{Y}\in\cm_{i}^{\ell}$. Let $\bm{X}\in\cb_{\srk,\ell}\left(\bm{0},r\right)$. Then $d_{\srk,\ell}\left(\bm{X},\bm{0}\right)\le r$, and hence by Proposition \ref{prop:2} we know that $d_{\srk,\ell}\left(\bm{X}\pm \bm{Y},\pm \bm{Y}\right)\le r$, which implies that $\bm{X}\pm \bm{Y}\in\cb_{\srk,\ell}\left(\pm \bm{Y},r\right)$. Conversely, let $\bm{X}\pm \bm{Y}\in\cb_{\srk,\ell}\left(\pm \bm{Y},r\right)$. Then $d_{\srk,\ell}\left(\bm{X}\pm \bm{Y},\pm \bm{Y}\right)\le r$, and hence by Proposition \ref{prop:2} we know that $d_{\srk,\ell}\left(\bm{X},\bm{0}\right)\le r$, which implies that $\bm{X}\in\cb_{\srk,\ell}\left(\bm{0},r\right)$. Let $\alpha\in\mathbb{F}_{q}\setminus\{0_{\mathbb{F}_q}\}$. Let $\bm{X}\in\cb_{\srk,\ell}\left(\bm{0},r\right)$. Then by Proposition \ref{prop:2} we have $d_{\srk,\ell}\left(\bm{X},\bm{0}\right)=d_{\srk,\ell}\left(\alpha \bm{X},\bm{0}\right)\le r$, which implies that $\alpha \bm{X}\in\cb_{\srk,\ell}\left(\bm{0},r\right)$. Conversely, let $\alpha \bm{X}\in\cb_{\srk,\ell}\left(\bm{0},r\right)$. Then by Proposition \ref{prop:2} again we have $d_{\srk,\ell}\left(\alpha \bm{X},\bm{0}\right)=d_{\srk,\ell}\left(\bm{X},\bm{0}\right)\le r$, which implies that $\bm{X}\in\cb_{\srk,\ell}\left(\bm{0},r\right)$.
\hfill$\square$
\end{pro}

\begin{pro}[Proposition \ref{prop:1}]
Let $r\in\left[0:\ell\min\left\{m,\eta\right\}\right]$. Since $\min\left\{m,\eta\right\}<\left(m+\eta\right)/2$ for all $m,\eta\in\mathbb{N}$, it is \textit{impossible} that $r\ge\ell\left(m+\eta\right)/2$.
\begin{enumerate}
\item Upper bound. By \cref{eq:SRBallsize} and the upper bound of $|\cs_{\srk,\ell}\left(\bm{0},r\right)|$ given in \cref{eq:4} we have:
we obtain:
\begin{equation*}
\left|\cb_{\srk,\ell}\left(\bm{0},r\right)\right|\le \sum_{s=0}^{r}K_{q}^{-\ell}\binom{\ell+s-1}{\ell-1}q^{\left(m+\eta-s/\ell\right)s}.
\end{equation*}
It is clearly that the binomial coefficient $C_{\ell+s-1}^{\ell-1}=\left(\ell+s-1\right)!/(\left(\ell-1\right)!s!)$ monotonically increases with $s$, and the function $f\left(s\right):=\left(m+\eta-s/\ell\right)s$ attains its maximum at $s=\ell\left(m+\eta\right)/2$. However, since $r<\ell\left(m+\eta\right)/2$ for all $r\in\left[0,\ell\min\left\{m,\eta\right\}\right]$, the maximum of $f\left(s\right)$ is $f\left(r\right)=q^{\left(m+\eta-r/\ell\right)r}$ for all $r\in\left[0:\ell\min\left\{m,\eta\right\}\right]$ and $s\in\left[0:r\right]$. Then:
\begin{equation*}
\begin{array}{ll}
\displaystyle{\sum_{s=0}^{r}K_{q}^{-\ell}\binom{\ell+s-1}{\ell-1}q^{\left(m+\eta-s/\ell\right)s}} & \displaystyle{\le\sum_{s=0}^{r}K_{q}^{-\ell}\binom{\ell+s-1}{\ell-1}\max_{s\in\left[0:r\right]}\left\{q^{\left(m+\eta-s/\ell\right)s}\right\}} \vspace{1ex}\\
\displaystyle{} & \displaystyle{=K_{q}^{-\ell}q^{\left(m+\eta-r/\ell\right)r}\sum_{s=0}^{r}\binom{\ell+s-1}{\ell-1}.} 
\end{array}
\end{equation*}
Recall the relative Fermat's equality:
\begin{equation*}
\sum_{\alpha=0}^{\gamma}\binom{\alpha+\beta}{\alpha}=\binom{\beta+\gamma+1}{\gamma}.
\end{equation*}
By taking $\alpha=s$, $\beta=\ell-1$, and $\gamma=r$ we obtain:
\begin{equation*}
K_{q}^{-\ell}q^{\left(m+\eta-r/\ell\right)r}\sum_{s=0}^{r}\binom{\ell+s-1}{\ell-1}=K_{q}^{-\ell}q^{\left(m+\eta-r/\ell\right)r}\binom{\ell-1+r+1}{r},
\end{equation*}
that is:
\begin{equation*}
\left|\cb_{\srk,\ell}\left(\bm{0},r\right)\right|\le K_{q}^{-\ell}q^{\left(m+\eta-r/\ell\right)r}\binom{\ell+r}{r}=K_{q}^{-\ell}q^{\left(m+\eta-r/\ell\right)r}\binom{\ell+r}{\ell}.
\end{equation*}
\item Lower bound. The lower bound is trivial\footnote{See also \cite[Lemma 2]{Ott2021}.}:
\begin{equation*}
\left|\cb_{\srk,\ell}\left(\bm{0},r\right)\right|=\sum_{s=0}^{r}\left|\cs_{\srk,\ell}\left(\bm{0},s\right)\right|\ge \left|\cs_{\srk,\ell}\left(\bm{0},r\right)\right|\ge K_{q}^{\ell}q^{\left(m+\eta-r/\ell\right)r-\ell/4}.
\end{equation*}
\hfill$\square$
\end{enumerate}
\end{pro}

\begin{pro}[Lemma \ref{lem:8}]
Let $a,b\in\mathbb{N}$ and $\alpha\in\left(0,1\right)$. Suppose that $\alpha ab\in\mathbb{N}$. Then $\log_q(C_{b}^{\alpha ab+b})\le b\log_q\left(\alpha ab+b\right)$. Let $l_ak\le a\le u_{a}k$ and $l_bk\le b\le u_b k$ for some $k\in\mathbb{N}$ and $l_a,l_b,u_a,u_b>0$. Then $b\log_q\left(\alpha ab+b\right)\le u_bk\log_q(\alpha u_au_bk^2+u_bk)=u_bk\log_q(\alpha u_ak+1)+u_bk\log_q(u_bk)$. Thus, for sufficiently large $k$ the term $\log_q(C_{b}^{\alpha ab+b})$ is at most of quasilinear order.
\hfill$\square$
\end{pro}

\begin{pro}[Lemma \ref{lem:2}]
Let $\eta$, $\ell$, and $w$ be positive integers. It is enough to show the case for $\ell=2$. The general case can be proven inductively. For $\ell=2$ we have to show that:
\begin{equation*}
\begin{bmatrix}
2\eta \\
w
\end{bmatrix}_q\ge\sum_{\bm{w}\in\ccp_2\left(w\right)}\begin{bmatrix}
\eta \\
w_1
\end{bmatrix}_q\begin{bmatrix}
\eta \\
w_2
\end{bmatrix}_q
\end{equation*}
By using the $q$-Vandermonde identity \cite{Sta2011} we have:
\begin{equation*}
\begin{bmatrix}
2\eta \\ w
\end{bmatrix}_q=\sum_{w_1=0}^{w}q^{w_1\left(w_1+\eta-w\right)}\begin{bmatrix}
\eta \\ w_1
\end{bmatrix}_q\begin{bmatrix}
\eta \\ w-w_1
\end{bmatrix}_q.
\end{equation*}
By defining $w_2:=w-w_1$ we obtain:
\begin{equation*}
\begin{bmatrix}
2\eta \\ w
\end{bmatrix}_q=\sum_{\bm{w}\in\ccp_2\left(w\right)}q^{w_1\left(\eta-w_2\right)}\begin{bmatrix}
\eta \\ w_1
\end{bmatrix}_q\begin{bmatrix}
\eta \\ w_2
\end{bmatrix}_q\ge\sum_{\bm{w}\in\ccp_2\left(w\right)}\begin{bmatrix}
\eta \\ w_1
\end{bmatrix}_q\begin{bmatrix}
\eta \\ w_2
\end{bmatrix}_q
\end{equation*}
for all $q\ge 2$.
\hfill$\square$
\end{pro}
\begin{pro}[Lemma \ref{lem:3}]
\upshape
Let $\cv$ be an $\mathbb{F}_q$-vector space of dimension $\eta$.
\begin{enumerate}
\item Note that:
\begin{equation}\label{eq:34}
\left|\cd_{\ell}\left(w,\cv^{\ell}\right)\right|=\sum_{\bm{w}\in\ccp_{\ell}\left(w\right)}\prod_{i=1}^{\ell}\begin{bmatrix}
\eta \\ w_i
\end{bmatrix}_q{\ge}\max_{\bm{w}\in\ccp_{\ell}\left(w\right)}\left\{\prod_{i=1}^{\ell}q^{\left(\eta-w_i\right)w_i}\right\}{=}:q^{\eta w}q^{\theta_{\ast}},
\end{equation}
where:
\begin{equation}\label{eq:35}
\theta_{\ast}:=\max_{\bm{w}\in\ccp_{\ell}\left(w\right)}\left\{-\sum_{i=1}^{\ell}w_i^2\right\}.
\end{equation}
To find $\theta_{\ast}$ we have to solve the optimization problem in \cref{eq:35} by using Lagrange multiplier. The optimization problem in \cref{eq:35} can be rewritten as the following standard form:
\begin{equation}\label{eq:36}
\begin{array}{lll}
\displaystyle{\max} & \displaystyle{-\sum_{i=1}^{\ell}w_i^2} & \displaystyle{} \vspace{1ex}\\
\displaystyle{\mathrm{subject\,\,to}} & \displaystyle{w_1+w_2+\ldots+w_{\ell}-w=0} & \displaystyle{} \vspace{1ex}\\
\displaystyle{} & \displaystyle{w_i-t_i^2=0,} & \displaystyle{i\in\left[1:\ell\right]} \vspace{1ex}\\
\displaystyle{} & \displaystyle{w_i-\eta+s_i^2=0,} & \displaystyle{i\in\left[1:\ell\right],}
\end{array}
\end{equation}
wherer $s_i$ and $t_i$ are $2\ell$ slack variables for all $i\in\left[1:\ell\right]$. The Lagrangian of the optimization problem in \cref{eq:36} is given by:
\begin{equation*}
\begin{array}{ll}
\displaystyle{} & \displaystyle{L\left(w_1,w_2,\ldots,w_\ell;\lambda,\mu_1,\mu_2,\ldots,\mu_\ell,\nu_1,\nu_2,\ldots,\nu_\ell\right)} \vspace{1ex}\\
\displaystyle{:=} & \displaystyle{-\sum_{i=1}^{\ell}w_i^2+\lambda\left(\sum_{i=1}^{\ell}w_i-w\right)+\sum_{j=1}^{\ell}\mu_j\left(w_j-t_j^2\right)+\sum_{l=1}^{\ell}\nu_{l}\left(w_l-\eta+s_l^2\right),}
\end{array}
\end{equation*}
Then we compute partial derivatives with respect to $w_1,w_2,\ldots,w_{\ell}$ and $\lambda,\mu_1,\mu_2,\ldots,\mu_{\ell},\nu_1,\nu_2,\ldots,\nu_{\ell}$, and set them to $0$ (the necessary conditions for an optimum):
\begin{equation*}
\begin{array}{ll}
\displaystyle{\partial_{w_i}L=-2w_i+\lambda+\mu_i+\nu_i\overset{!}{=}0,} & \displaystyle{i\in\left[1:\ell\right]} \vspace{1ex}\\
\displaystyle{\partial_{\lambda}L=w_1+w_2+\ldots+w_\ell-w\overset{!}{=}0} & \displaystyle{} \vspace{1ex}\\
\displaystyle{\partial_{\mu_i}L=w_i-t_i^2\overset{!}{=}0,} & \displaystyle{i\in\left[1:\ell\right]} \vspace{1ex}\\
\displaystyle{\partial_{t_i}L=-2\mu_it_i\overset{!}{=}0,} & \displaystyle{i\in\left[1:\ell\right]} \vspace{1ex}\\
\displaystyle{\partial_{\nu_i}L=w_i-\eta+s_i^2\overset{!}{=}0,} & \displaystyle{i\in\left[1:\ell\right]} \vspace{1ex}\\
\displaystyle{\partial_{s_i}L=2\nu_is_i\overset{!}{=}0,} & \displaystyle{i\in\left[1:\ell\right],} 
\end{array}
\end{equation*}
where $\partial_{\cdot}$ is the shorthand for $\partial/\partial\cdot$. From $-2\mu_it_i^2=0$ and $2\nu_is_i=0$ for each $i\in\left[1:\ell\right]$ we set $\mu_i=\nu_i=0$ for each $i\in\left[1:\ell\right]$. Substituting $\mu_i=\nu_i=0$ into $-2w_i+\lambda+\mu_i+\nu_i=0$ yields $w_i=\lambda/2$ for all $i\in\left[1:\ell\right]$. Further, from $w_1+w_2+\ldots+w_{\ell}-w=0$ we obtain $\lambda=2w/\ell$ and hence $w_i=w/\ell$ for all $i\in\left[1:\ell\right]$. We have to check if this solution is feasiable. For each $i\in\left[1:\ell\right]$ substituting $w_i=w/\ell$ into $w_i-t_i^2=0$ and $w_i-\eta+s_i^2=0$ yields $t_i^2=w/\ell$ and $s_i^2=\eta-w/\ell$. Clearly, $t_i^2>0$ for all $i\in\left[1:\ell\right]$. Actually we have $s_i^2\ge 0$ for all $i\in\left[1:\ell\right]$ as well, since $w_i\le \eta$ for each $i\in\left[1:\ell\right]$ and hence $\eta-w/\ell=\eta-\left(w_1+w_2+\ldots+w_\ell\right)/\ell\ge \eta-\eta\ell/\ell=0$. Thus, the solution $w_i=w/\ell$ for each $i\in\left[1:\ell\right]$ is feasible. Therefore:
\begin{equation*}
\theta_{\ast}=-\sum_{i=1}^{\ell}\left(\frac{w}{\ell}\right)^2=-\frac{w^2}{\ell}.
\end{equation*}
Finally, substituting $\theta_{\ast}=-w^2/\ell$ into \cref{eq:34} yields $\left|\cd_{\ell}\left(w,\cv^{\ell}\right)\right|\ge q^{\eta w-w^2/\ell}$, which is the desired lower bound. To derive the upper bound, we use the following estimation:
\begin{equation*}
\begin{array}{ll}
\displaystyle{\left|\cd_{\ell}\left(w,\cv^{\ell}\right)\right|} & \displaystyle{=\sum_{\bm{w}\in\ccp_{\ell}\left(w\right)}\prod_{i=1}^{\ell}\begin{bmatrix}
\eta \\ w_i
\end{bmatrix}_q\le\left|\ccp_\ell\left(w\right)\right|\max_{\bm{w}\in\ccp_{\ell}\left(w\right)}\left\{\prod_{i=1}^{\ell}K_q^{-1}q^{\left(\eta-w_i\right)w_i}\right\}} \vspace{1ex}\\
\displaystyle{} & \displaystyle{\le K_q^{-\ell}\binom{w+\ell-1}{\ell-1}q^{\eta w-w^2/\ell}.}
\end{array}
\end{equation*}
\item Let $\cx:=\Pi_{i}^{\ell}\cx_{i}$ and $\cy:=\Pi_{i}^{\ell}\cy_{i}$ be two $\ell$-decomposable subspaces of $\mathbb{F}^{\eta\ell}_q$. Let $\bm{v}:=\left(v_1,\ldots,v_{\ell}\right)\in\left(\cx\cap\cy\right)$. Then $\bm{v}\in\Pi_{i}^{\ell}\cx_{i}$ and $\bm{v}\in\Pi_{i}^{\ell}\cy_{i}$, which imply that $v_i\in\left(\cx_i\cap\cy_i\right)$ for all $i\in\left[1:\ell\right]$ and hence $\bm{v}\in\Pi_{i}^{\ell}\left(\cx_i\cap\cy_i\right)$. Conversely, let $\bm{v}\in\Pi_{i}^{\ell}\left(\cx_i\cap\cy_i\right)$. Then $v_i\in\cx_i$ and $v_i\in\cy_i$ for all $i\in\left[1:\ell\right]$, which imply that $\bm{v}\in\left(\cx\cap\cy\right)$. Thus, we proved that $\cx\cap\cy=(\Pi_{i}^{\ell}\cx_{i})\cap(\Pi_{i}^{\ell}\cy_i)=\Pi_{i}^{\ell}\left(\cx_i\cap\cy_i\right)$ and hence $\dim\left(\cx\cap\cy\right)=\sum_{i\in\left[1:\ell\right]}\dim\left(\cx_i\cap\cy_i\right)$.
\item Let $\cx_i,\cy_i\le\cv$ for all $i\in\left[1:\ell\right]$. Then $\Pi_{i}^{\ell}\cx_i$ and $\Pi_{i}^{\ell}\cy_i$ are subspaces of $\cv^{\ell}$. Thus:
\begin{equation*}
\Pi_{i}^{\ell}\cx_i+\Pi_{i}^{\ell}\cy_i=\left\{\left(x_1+y_1,x_2+y_2,\ldots,x_{\ell}+y_{\ell}\right):x_i\in\cx_i,y_i\in\cy_i,i\in\left[1:\ell\right]\right\},
\end{equation*}
is a subspace of $\cv^{\ell}$. Moreover, since $\cx_i$ and $\cy_i$ are subspaces of $\cv$ for all $i\in\left[1:\ell\right]$, then sums $\cx_i+\cy_i$ are subspaces of $\cv$, and hence $\Pi_{i}^{\ell}\left(\cx_i+\cy_i\right)$ is a subspace of $\cv^{\ell}$, which is of the form:
\begin{equation*}
\Pi_{i}^{\ell}\left(\cx_i+\cy_i\right)=\left\{\left(x_1+y_1,x_2+y_2,\ldots,x_{\ell}+y_{\ell}\right):x_i\in\cx_i,y_i\in\cy_i,i\in\left[1:\ell\right]\right\}.
\end{equation*}
In summary, we have proven that $\Pi_{i}^{\ell}\cx_i+\Pi_{i}^{\ell}\cy_i$ and $\Pi_{i}^{\ell}\left(\cx_i+\cy_i\right)$ are two subspaces of $\cv^{\ell}$, and that they contain exactly the same elements. Thus, we have $\Pi_{i}^{\ell}\cx_i+\Pi_{i}^{\ell}\cy_i=\Pi_{i}^{\ell}\left(\cx_i+\cy_i\right)$. The dimension of $\cx+\cy$ can be determined by the well-known dimension formula:
\begin{equation*}
\begin{array}{ll}
\displaystyle{\dim\left(\cx+\cy\right)} & \displaystyle{=\dim\left(\Pi_{i}^{\ell}\left(\cx_i+\cy_i\right)\right)=\sum_{i=1}^{\ell}\dim\left(\cx_i+\cy_i\right)} \vspace{1ex}\\
\displaystyle{} & \displaystyle{=\sum_{i=1}^{\ell}\left(\dim\left(\cx_i\right)+\dim\left(\cy_i\right)-\dim\left(\cx_i\cap\cy_i\right)\right)} \vspace{1ex}\\
\displaystyle{} & \displaystyle{=w_{x}+w_{y}-\sum_{i=1}^{\ell}\dim\left(\cx_i\cap\cy_i\right).}
\end{array}
\end{equation*}
\hfill$\square$
\end{enumerate}
\end{pro}

\newpage
\bibliography{refs}
\bibliographystyle{alpha}
\end{document}